\documentclass[%
reprint,
superscriptaddress,
 amsmath,amssymb,
 aps,
 pre,
floatfix,
]{revtex4-1}

\usepackage{graphicx}
\usepackage{dcolumn}
\usepackage{bm}

\usepackage{epstopdf}
\usepackage{lipsum}
\usepackage[]{units}
\usepackage{upgreek}
\usepackage{mathrsfs}

\usepackage[colorlinks=true,citecolor=blue]{hyperref}
\usepackage{natbib}

\begin{document}

\title{Prospects for laser-driven ion acceleration through controlled displacement of electrons by standing waves}

\author{J. Magnusson}
\email{joel.magnusson@chalmers.se}
\affiliation{Department of Physics, Chalmers University of Technology, SE-412 96 Gothenburg, Sweden}

\author{F. Mackenroth}
\affiliation{Department of Physics, Chalmers University of Technology, SE-412 96 Gothenburg, Sweden}
\affiliation{Max Planck Institute for the Physics of Complex Systems, N\"othnitzer Str. 38, 01187 Dresden, Germany}

\author{M. Marklund}
\affiliation{Department of Physics, Chalmers University of Technology, SE-412 96 Gothenburg, Sweden}

\author{A. Gonoskov}
\affiliation{Department of Physics, Chalmers University of Technology, SE-412 96 Gothenburg, Sweden}
\affiliation{Institute of Applied Physics, Russian Academy of Sciences, Nizhny Novgorod 603950, Russia}
\affiliation{Lobachevsky State University of Nizhni Novgorod, Nizhny Novgorod 603950, Russia}

\date{\today} 

\begin{abstract}
During the interaction of intense femtosecond laser pulses with various targets, the natural mechanisms of laser energy transformation inherently lack temporal control and thus commonly do not provide opportunities for a controlled generation of a well-collimated, high-charge beam of ions with a given energy of particular interest. In an effort to alleviate this problem, it was recently proposed that the ions can be dragged by an electron bunch trapped in a controllably moving potential well formed by laser radiation. Such standing-wave acceleration (SWA) can be achieved through reflection of a chirped laser pulse from a mirror, which has been formulated as the concept of chirped-standing-wave acceleration (CSWA). Here we analyze general feasibility aspects of the SWA approach and demonstrate its reasonable robustness against field structure imperfections, such as those caused by misalignment, ellipticity and limited contrast. Using this we also identify prospects and limitations of the CSWA concept.
\end{abstract}

\pacs{Valid PACS appear here} 


\maketitle


\section{Introduction}
The generation of high-energy ions via the interaction of high-intensity femtosecond laser pulses with various targets provides a promising basis for a new kind of compact ion sources with numerous applications in medicine, industry and science \cite{Daido,Macchi,Bulanov}. Over the last couple of decades, extensive theoretical and experimental studies have made it possible to reveal and understand the natural mechanisms of energy transformation from laser radiation to kinetic energy of ions. This has further made it possible to identify several favourable interaction regimes and to develop related concepts, including target normal sheath acceleration (TNSA) \cite{hatchett.pop.2000, wilks.pop.2001, mackinnon.prl.2002, Roth,Mora,Cowan,Passoni1}, Coulomb explosion (CE) of clusters \cite{Ditmire,Kovalev1,Kovalev2}, double-layered targets \cite{Esirkepov1,Esirkepov2,Bulanov2}, breakout afterburner (BOA) \cite{yin.lpb.2006,yin.pop.2007}, hole boring \cite{Schlegel}, collisionless shock acceleration \cite{Silva,Haberberger}, magnetic vortex acceleration \cite{bulanov.ppr.2005,nakamura.prl.2010} and light sail or radiation pressure acceleration \cite{Esirkepov3,Bulanov3,Henig,Kar}.

Due to their natural robustness the most experimentally accessible schemes are based on plasma heating as the first stage of energy transformation. A number of studies have recently been performed on specially designed targets and laser pulse shapes \cite{Flippo,Buffechoux,Burza,Gaillard,Markey,Pfotenhauer,Dalui,Floquet,Jiang,Margarone1,Passoni2,Zou} as well as nano- and microstructured targets \cite{Lubcke,Nodera,Cao,Klimo,Margarone2,Andreev1,Blanco,Andreev2,Magnusson} in order to enhance the energy coupling and thus increase overall efficiency of both TNSA and CE. Although increasing the achievable energy of ions is of crucial importance, some recent studies are also focused on achieving high flux \cite{mackenroth.epjd.2017} and enhancing or controlling collimation \cite{kar.ncom.2016, gonzalez.ncom.2016, giuffrida.prab.2017} of the ion beams so that they can meet the requirements of particular applications. In this respect, despite being accessible and sufficient for some applications, the natural mechanisms have intrinsic limitations that preclude meeting the requirements of more advanced applications. One of the fundamental reasons behind this is a lack of temporal control over the processes, therefore providing us with no advanced means for an controlled conversion of laser energy into kinetic energy of ions moving in a chosen direction with given energy. 

An interesting approach to the creation of a controllable acceleration process has recently been proposed in Ref. \cite{Mackenroth}. In this paper the authors demonstrated numerically that proton bunches with energies of \unit[100]{MeV} can be produced in a controllable manner using a \unit[30]{J} laser pulse. Their approach implies dragging ions with electrons that are gradually shifted while being locked by a laser-formed standing wave. To gradually change the position of the locked electrons, and thereby continuously accelerate the ions, the authors of Ref. \cite{Mackenroth} proposed to use a chirped laser pulse and thus named their concept Chirped-Standing-Wave Acceleration (CSWA). However, other ways of controlling the position of such locked electrons are likely to be developed in the future (see, for example, Ref. \cite{Wan}) and may provide more advanced and efficient ways of using this general approach, which we refer to as Standing-Wave Acceleration (SWA). In this paper we thus consider SWA separately as a basic promising approach for laser-driven ion acceleration. We assess the feasibility of and prospects for the implementation of this approach in future experiments. We also assess the prospects for and limitations of performing a proof-of-principle demonstration of SWA based on the CSWA concept.

\vfill

\section{Controlling the acceleration}

\begin{figure}[!t]
  \includegraphics[width=\columnwidth]{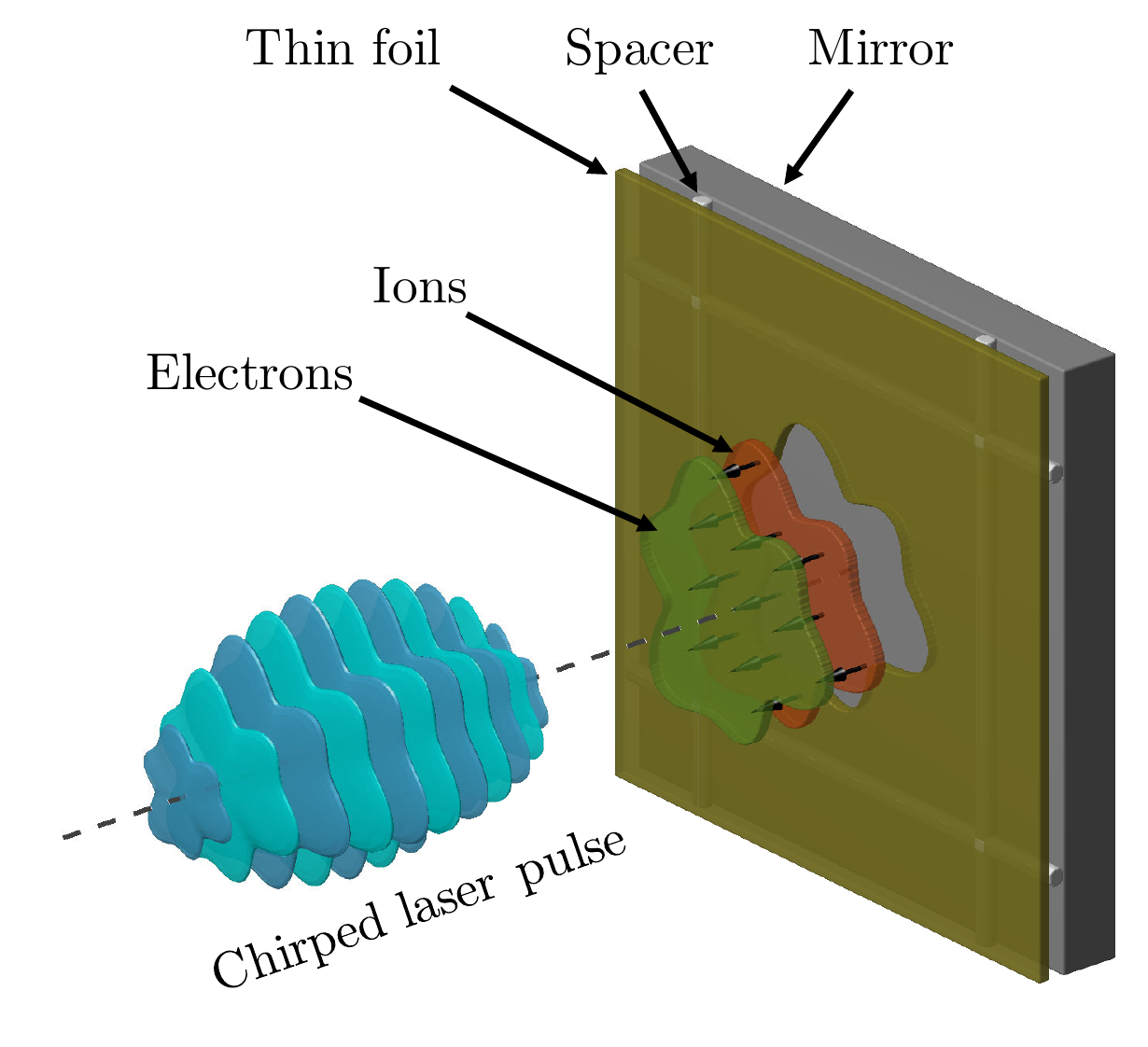}
\caption{Schematic representation of SWA's capability for controlling the acceleration process, as it appears in the CSWA implementation. It shows the prospects of generating monoenergetic ions in spite of having a complex transverse intensity profile of the laser pulse. The target consists of a thin foil of electrons and ions to be accelerated, suspended on top of a heavy mirror by a micron-sized spacer. The ions are accelerated by an electrostatic field, generated when the electrons are trapped and pulled away by the standing wave formed when reflecting the laser field from the mirror. This acceleration occurs within an area where the laser intensity exceeds the RSIT threshold and the shape of this area therefore matches that of the laser pulse.}
\label{fig:setup3D}
\end{figure}

When developing advanced approaches, it is important to keep in mind the experimental feasibility of these approaches. For example, super-Gaussian or donut shaped transverse distributions for the driving laser pulse can help to control transverse instabilities and/or improve collimation and monoenergecity by mitigating the dependency of the acceleration on the intensity variation in the transverse direction. However, our ability to accurately control the laser pulse shape in experiments is limited and the implementation of such ideas might therefore require challenging technical developments. 

While the exact motion of charged particles in an intense laser field can be non-trivial, relativistic self-induced transparency (RSIT) arises naturally as a consequence of the speed limit that is the speed of light. In order for a plasma to be opaque to the incoming laser field it must be able to generate a field that can cancel out any otherwise transmitted radiation. However, for a given number of charged particles there is a maximal current that can be sustained, since all particles are restricted to speeds less than the speed of light. This also means that there is a maximal field strength that the particles can generate and they will as a result be unable to cancel out laser fields surpassing this threshold, in effect becoming transparent to them.

In the majority of schemes, the dependency of the ion acceleration rate on the local intensity precludes generation of quasi-monoenergetic ion beams because the intensity inevitably varies along both the longitudinal and traverse directions. The idea of using the threshold effect of RSIT for a sequence of thin separated films as means of accelerating ions controllably through electron displacement by the ponderomotive force in a sequentially-triggered manner has been proposed in Ref.~\cite{gonoskov.prl.2009}. In this case the acceleration process does not primarily depend on laser intensity but is merely enabled as the transparency threshold is surpassed. In such a case, the acceleration rate will instead be proportional to the areal density of the plasma generated from the film. Thus, for a given area in the transverse plane, ions are accelerated along the direction of pulse propagation at a fixed rate, independently on the distance to the propagation axis of the pulse. The duration of the acceleration, however, depends on the distance to this axis because the transparency occurs earlier and for a longer period on axis than at the periphery. 

\begin{figure}[!t]
  \includegraphics[width=\columnwidth]{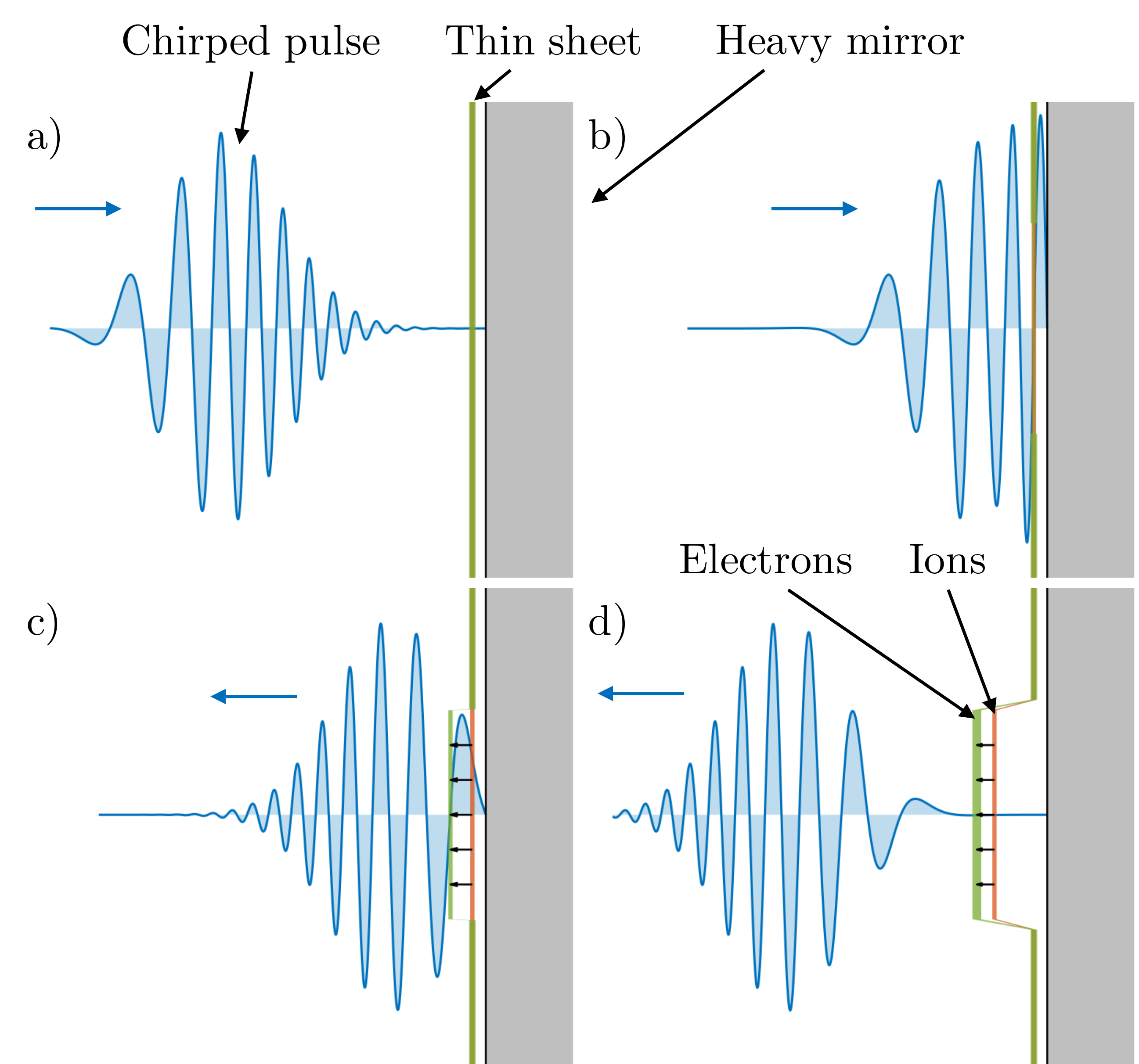}
\caption{Schematic representation of the general scenario of the CSWA concept. (a) A chirped laser pulse incident on a high-density mirror with a thin overdense layer fixed in a position some distance from the mirror. (b) The pulse penetrates the thin layer as it becomes relativistically transparent, forming a standing wave which compresses and locks the electrons to the electric field node. (c) As the frequency of the standing wave decreases the field nodes move away from the mirror and the locked electrons are consequently pulled along, setting up an electrostatic field between them and protons of the thin layer. (d) The electrons are released as the pulse leaves the mirror and the standing wave disappears. The protons, having obtained a significant amount of energy, is travelling away from the mirror.}
\label{fig:setup}
\end{figure}

Disentangling the acceleration time from the local intensity of radiation was achieved in the recently proposed concept of chirped-standing-wave acceleration (CSWA) \cite{Mackenroth}. In this concept, a circularly polarized chirped laser pulse is normally incident on a thin film which is placed, at some small distance, in front of a thick solid target (see Figure \ref{fig:setup3D} and \ref{fig:setup}). During the first stage, the laser radiation surpasses the RSIT threshold for the thin layer and quickly forms a standing wave as the pulse is reflected from the thick target, acting as a mirror. If the thin layer is placed at the position of one of the nodes of this standing wave, the electrons of the thin layer become locked from both sides by the ponderomotive force. For the discussion we will consider only the first node, which occurs at half the wavelength of the laser radiation in the foregoing part of the laser pulse. The areal density of the thin foil is assumed to be sufficiently low, such that the fields produced by the locked electrons cannot significantly perturb the standing wave. Under such conditions, the locked electrons are well localized and predominantly just rotate in the external field of the standing wave and their localization is therefore not subject to instabilities in the transverse direction. 

Since the electric field node initially coincides with the initial position of the thin layer the electrons will not be shifted relative to the ions in the longitudinal direction, and no ion acceleration will therefore yet occur. In this way, prior to any ion acceleration, we can lock electrons within an area in the transverse plane, where the intensity during the first stage surpasses the relativistic transparency threshold. This \textit{acceleration area} can have an arbitrary shape, see Figure \ref{fig:setup3D}, and is only required to have a typical size that is large compared to the laser wavelength.

The acceleration is enabled and controlled by varying the wavelength, which is one of the most well- and accurately-controlled parameters in laser technologies. In the simplest implementation of the CSWA concept, the chirp of the laser pulse provides a gradual shift of the node position relative to the mirror which results in an accurate and gradual displacement of the locked electrons, relative to the ions of the thin layer. Since the areal density of the locked electrons is the same everywhere within the acceleration area, their displacement gives rise to the formation of a microscopic capacitor with uniform and unidirectional longitudinal electric field. Thus, within the acceleration area the acceleration rate of the ions by this field does not depend on their lateral position. Certainly, the strength of this accelerating field depends on the position of the ion in the longitudinal direction. However, if the motion of the node is sufficiently slow, the fastest ions will overtake the electrons and then be decelerated. Thus, by adjusting the node velocity over time we can cause the ions to oscillate around the locked electrons, accompanied by a gradual acceleration, such that quasi-monoenergicity is maintained. In such a way, tuning the chirp and adjusting the areal density of the layer makes it possible to control the number of accelerated ions, their average energy and energy spread. It should also be noted that the idea of forming the standing wave by reflecting the laser pulse from a mirror automatically implies perfect spatial and temporal overlap. In addition to this the longitudinal position of the locked electrons, and consequently the acceleration process, does not depend on the phase of the laser radiation.

The fact that the acceleration process is essentially insensitive to the laser pulse shape, intensity and lateral phase distribution, suggests that the idea of ion acceleration by locked electrons can be considered as an essential basis for time-controlled acceleration. From the qualitative analysis provided thus far, we therefore conclude that this idea has the potential of producing high-quality, well-collimated, quasi-monoenergetic beams of ions. In the interest of providing a more accurate, quantitative analysis, we in this paper assess the tolerance of this process against the factors most crucial for experimental implementation. 

First, a limited laser contrast naturally results in a thermal expansion of the plasma emerging from the ionization of the thin layer. We identify the acceptable range of the plasma expansion, before this starts to significantly affect the acceleration process. Next, since driving a laser pulse at normal incidence can lead to damage of the laser system by backreflection of the pulse, it is favourable to use oblique incidence in experiments. We here ascertain how large incidence angles are acceptable. Finally, the fact that the locked electrons are well localized is a result of the circular polarization of the laser field, something which linear polarization does not provide. We further identify the largest acceptable deviation from circular polarization. Apart from aspects of experimental feasibility, in this paper we also identify and discuss general prospects and limitations for CSWA as the basic implementation of the SWA approach.

We should also note that another interesting implementation of the SWA approach has recently been proposed in Ref.\,\cite{Wan}. In this work the authors propose using two counter-propagating pulses of different frequency and intensity to lock and move electrons of a thin layer. For this concept we cannot apply the foregoing arguments about the insensitivity to the alignment, intensity and lateral phase distribution. However, the identified tolerance to the angle of incidence, laser contrast and ellipticity is still relevant to this concept and thus supports its feasibility.

\section{Numerical setup}
The study is performed using PIC simulations carried out with the code \textsc{Picador} \cite{Bastrakov,Surmin}. The numerical setup consists of three parts: 1) a circularly polarized chirped laser pulse; 2) a dense thick foil acting as a reflecting mirror to the incoming laser radiation; and 3) a thin (sub-micron) sheet of protons and electrons positioned at some fixed distance from the mirror. The laser pulse has a Gaussian shape in both longitudinal and transverse directions. The model for the pulse chirp implies a retardation of harmonics by a distance linearly proportional to the frequency, such that the spectrum is preserved. The analytical expression for this model is described in Ref.\,\cite{Mackenroth}. 
Assuming that the laser radiation propagates along the $x$-axis of the rectangular coordinate system $xyz$, the longitudinal shape of the incoming laser field can be described by a function of the phase, $\eta = t - x/c$,
\begin{equation}\label{eq:ChirpedPulse}
  \Psi(\eta) =  \psi_c\exp{[-\alpha\eta^2]}\exp{[i(\omega_0\eta + \kappa\eta^2 + \delta)]}
\end{equation}
where $t$ is time, $c$ is the speed of light, $\omega_0$ is the laser central frequency and
\begin{equation}
  \begin{aligned}
    \psi_c = \frac{1}{\sqrt[4]{1 + \mathcal{C}^2}}, \quad 
    \alpha = \frac{1}{8\ln2(1+\mathcal{C}^2)}\Delta\omega^2, \\
    \kappa  = \mathcal{C}\alpha, \quad 
    \delta = \frac{2\ln2}{(\frac{\Delta\omega}{\omega_0})^2}\mathcal{C} + \frac{\arctan{\mathcal{C}}}{2}.
  \end{aligned}
\end{equation}
Here $\mathcal{C}$ is the dimensionless chirp parameter introduced in Ref.\,\cite{Mackenroth} and $\Delta\omega$ the laser FWHM bandwidth. The chirp parameter is related to the amount of stretching of the pulse by $\tau_c/\tau_0 = \sqrt{1+\mathcal{C}^2}$, where $\tau_0$ is the duration of the pulse when it is optimally compressed and $\tau_c$ when it is chirped. The duration of the unchirped pulse can in turn be related to the fractional bandwidth, which is the ratio between the bandwidth and the central frequency, which for a Gaussian pulse is
\begin{equation}\label{eq:duration}
\frac{\Delta\omega}{\omega_0} \approx 1.47 \left( \frac{\lambda_0}{\mu\mathrm{m}} \right) \left( \frac{\tau_0}{\mathrm{fs}} \right)^{-1},
\end{equation}
where $\tau_0$ is defined at FWHM and $\lambda_0 = 2\pi c/\omega_0 = \unit[810]{nm}$ is the central laser wavelength used throughout this study.

The longitudinal shape of the electric and magnetic fields are described by
\begin{equation}
\vec{E} = \frac{a_0}{\sqrt{2}}\Psi(\eta) [\hat y + i\hat z ] ,\quad 
\vec{B} = \hat{x} \times \vec{E},
\end{equation}
where $a_0 = eE_0/m_ec\omega_0$ is the normalized laser amplitude, $E_0$ the peak electric field strength, $m_e$ and $-e$ are the electron mass and charge, respectively. In the lateral directions the Gaussian shape of the laser pulse has the size $w$ (FWHM of intensity).

As the laser intensity surpasses the RSIT threshold the thin layer target becomes relativistically transparent, allowing the pulse to be transmitted. The radiation is then reflected from the heavy mirror, locally forming a standing wave with the first electric field node at a distance of half a wavelength from the mirror. The electrons are then compressed by the ponderomotive force of the standing wave and locked to the position of this node. As the pulse frequency changes over time, the field nodes are shifted in a direction determined by the sign of the chirp. Thus the electrons can in a controllable way be moved further away from the mirror, pulling the protons of the thin layer with them via the electrostatic field of charge separation. This is visualized in Figure \ref{fig:setup} and forms the general principles of the CSWA scheme.

As the ion dynamics is determined by the charge to mass ratio this can correspond to several different ion species at different levels of ionization. Increasing the charge to mass ratio by assuming a higher level of ionization would make the ions of the mirror more mobile, but it would also increase the density of free electrons in the mirror. With this in mind we here model the mirror as a plasma consisting of electrons and ions with a mass of $65m_p$ and charge $+e$, where $m_p$ is the proton mass, corresponding to for example triply ionized gold. The plasma density of the mirror $n_0$ is set at $150n_\mathrm{cr}$, where $n_\mathrm{cr} = m_e\omega_0^2/4\pi e^2$ is the critical plasma density. 

\begin{figure}[!t]
\centering
  \includegraphics[width=\columnwidth]{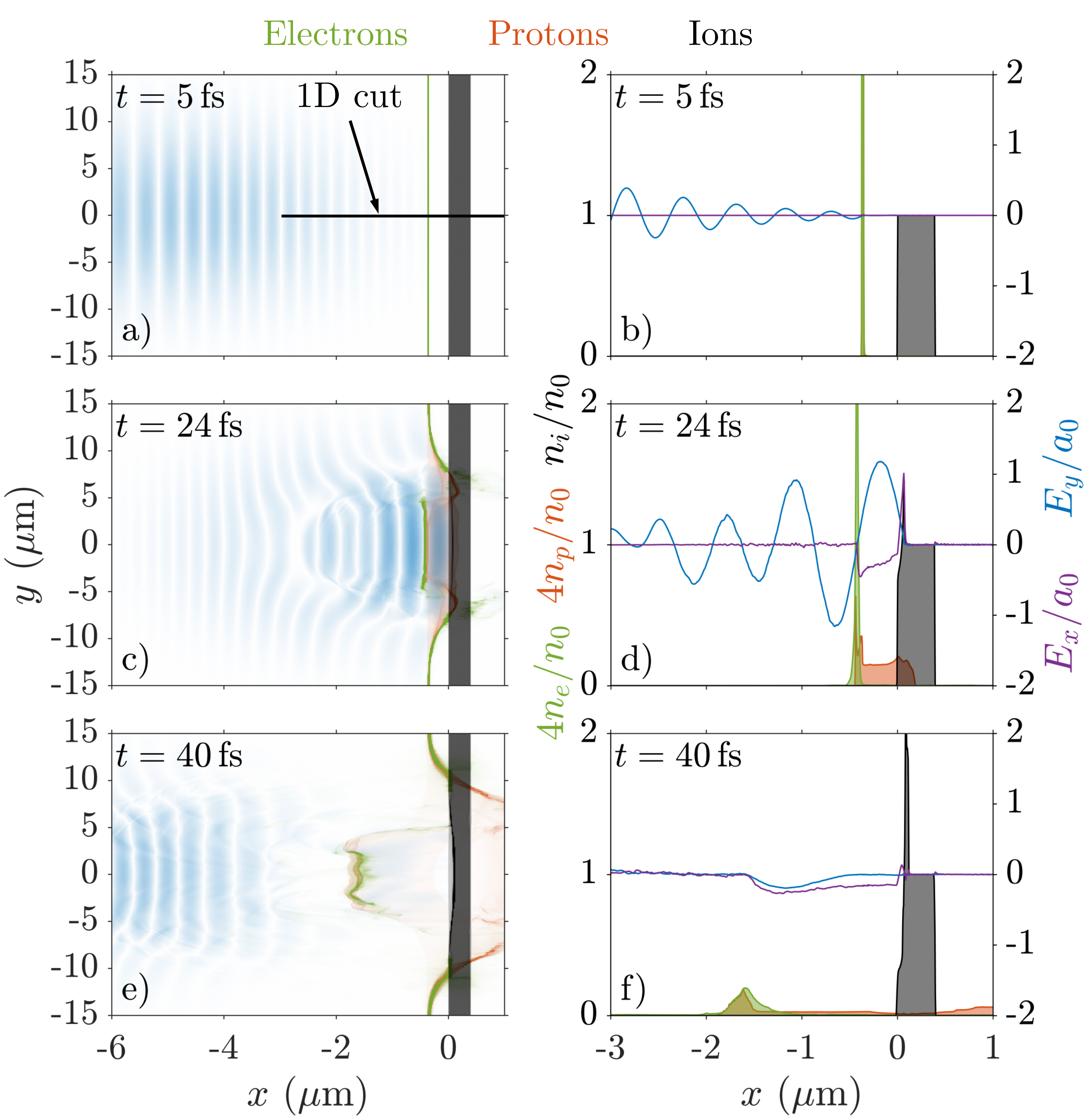}
\caption{A 2D PIC simulation for a laser energy $\varepsilon_0 = \unit[80]{J}$, bandwidth $\Delta\omega = 0.5\omega_0$ and chirp $\mathcal{C} = -4$ is shown for three time instants: before the interaction between the laser pulse and the thin foil (a)-(b); during the CSWA stage when the electrons are locked to the standing wave formed by the reflected radiation (c)-(d); and some time after the pulse has been reflected and the electrons released. (a), (c), (e) Magnitude of the transverse electric field $E_y$ (blue), electron density (green), proton density (red), and ion density (grey) as functions of 2D coordinates. (b), (d), (f) A 1D cut additionally showing the longitudinal electric field $E_x$ (purple) and transverse electric field $E_y$ (blue), with fields obtained for $y=0$ and densities averaged over the range $|y| < w/2$.}
\label{fig:ideal}
\end{figure}

To account for the slightly shorter instantaneous wavelengths at the leading edge of the chirped pulse the thin sheet is placed at distance of $0.45\lambda_0$ from the mirror, instead of $0.5\lambda_0$. It consists of protons and electrons with an areal density, $\sigma$, given in units of the critical areal density, which we define as $\sigma_\mathrm{cr} = n_\mathrm{cr}\lambda_0 = 2\pi c n_\mathrm{cr}/\omega_0$.

A CSWA simulation, performed under ideal conditions for later comparison, is shown in Figure \ref{fig:ideal} for several time instances both in the simulated 2D space and in a 1D cut along the pulse longitudinal direction. The simulation was carried out with laser energy $\varepsilon_0 = \unit[80]{J}$, bandwidth $\Delta\omega = 0.5\omega_0$ and chirp $\mathcal{C} = -4$. Furthermore, the pulse has a FWHM waist of $w = \unit[10]{{\mu}m}$ and thus a laser amplitude of $a_0 \approx 115$. The areal density of the thin sheet is $5\sigma_\mathrm{cr} = \unit[7\times 10^{17}]{cm^{-2}}$, corresponding to an ultra-thin ($\sim\unit[10]{nm}$) foil of solid density ($\sim\unit[10^{24}]{cm^{-3}}$).

Figure \ref{fig:ideal}(a-b) shows the simulation some time before the thin sheet becomes relativistically transparent. In Figure \ref{fig:ideal}(c-d) the electrons within the laser spot ($|y|<w/2$) are seen to be locked at the field node and has started to move in the negative $x$-direction and the electrostatic field, $E_x$, can now be clearly seen in Figure \ref{fig:ideal}(d). Furthermore, the ion distribution has now been shifted towards the mirror, as a result of their delayed response to the initial push excerted on the electrons. The majority of them will however get pulled back by the electrostatic field, starting when the electrons get locked to the field node. As the pulse passes, the protons continue to accelerate via a residual field locked between the thin sheet and the mirror, as seen in Figure \ref{fig:ideal}(e-f).

\section{The effect of limited contrast}\label{sec:thickness}
Realistic pulses are likely to have a non-negligible pedestal which would pre-heat the initially thin sheet, as well as the mirror, making it expand. In the analytical calculations performed so far the sheet has been modeled as being infinitely thin, which of course is not completely accurate. This assumption is however reasonable as long as the \emph{scale length}, $L$, of the expanded plasma of the thin sheet can be considered to be much smaller than the laser wavelength, $L \ll \lambda_0$, which is well within current capabilities. However, if the thin sheet was to expand to a scale length comparable to the laser wavelength, $L\sim\lambda_0$, the sheet dynamics can not reasonably be modeled as that of a thin sheet anymore. 

We here investigate the robustness of the CSWA scheme to the scale length of the thin sheet by performing 2D simulations, keeping the areal density of the sheet, $\sigma$, fixed at $5\sigma_\mathrm{cr}$. The sheet is initiated as a neutral electron-proton plasma with a density profile of an isosceles triangle and with the sheet scale length $L$ for its base, giving it a symmetric density up- and downramp.

\begin{figure}[!t]
\centering
  \includegraphics[width=\columnwidth]{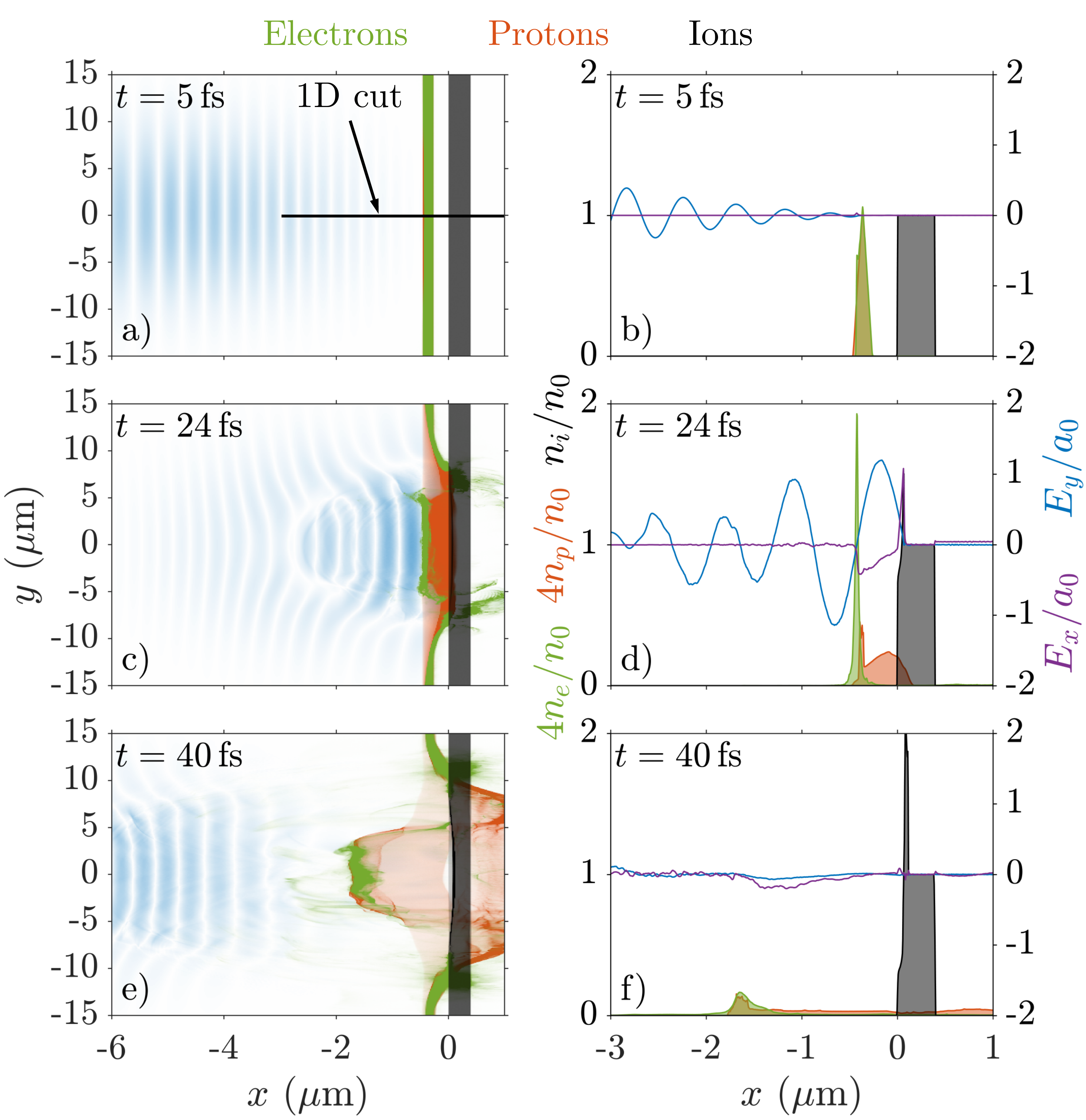}
\caption{A 2D PIC simulation for a laser energy $\varepsilon_0 = \unit[80]{J}$, bandwidth $\Delta\omega = 0.5\omega_0$ and chirp $\mathcal{C} = -4$. The sheet was initiated with a scale length $L = \lambda_0/4$. The information is presented as in Figure \ref{fig:ideal}.}
\label{fig:scalelength}
\end{figure}

In Figure \ref{fig:scalelength} the case of a sheet scale length of $\lambda_0/4$ is shown for several time instances both in the simulated 2D space and in a 1D cut along the pulse longitudinal direction. The simulations were performed for a circularly polarized pulse with laser parameters identical to that of the preceding section ($\varepsilon_0 = \unit[80]{J}$, $\Delta\omega = 0.5\omega_0$, $\mathcal{C} = -4$, $w = \unit[10]{{\mu}m}$).

Figure \ref{fig:scalelength}(a-b) shows the simulation some time before the relativistic transparency of the electron-proton sheet sets in and the initial density profile of the now relatively thick sheet can be clearly seen. In Figure \ref{fig:scalelength}(c-d) the electrons within the laser spot ($|y|<w/2$) are seen to be locked at the field node and has started to move in the negative $x$-direction. The electrostatic field, $E_x$, can be seen to have formed in Figure \ref{fig:scalelength}(d) and the compression of the electrons at the field node is also clearly visible by comparison with Figure \ref{fig:scalelength}(b). As the pulse passes, the protons continue to accelerate via a residual field locked between the thin sheet and the mirror, as seen in Figure \ref{fig:scalelength}(e-f). Comparing this to Figure \ref{fig:ideal} we see that they mainly only differ in that the residual field is much less pronounced as well as noisier when considering this expanded foil. Nevertheless, it clearly demonstrates the compression of the electrons at the field node and that the acceleration process is essentially unchanged.

\begin{figure}[!t]
  \includegraphics[width=\columnwidth]{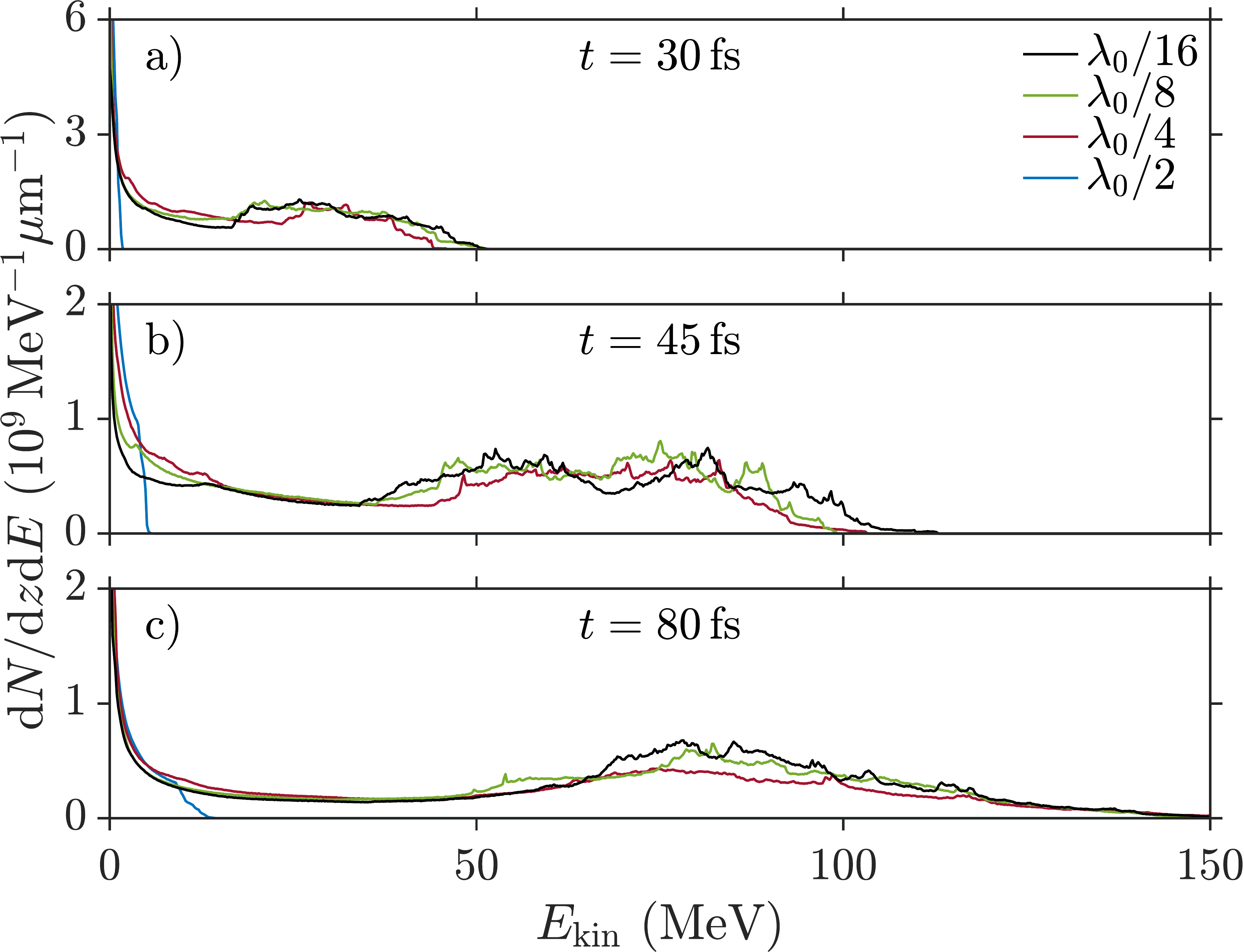}
\caption{Spectra of protons from the thin plasma layer for different sheet scale lengths, $L$,  shown at three times $\unit[30]{fs}$ (a), $\unit[45]{fs}$ (b) and $\unit[80]{fs}$ (c). Only protons within a distance of half a pulse waist from the $x$-axis ($|y| < w/2$) and moving in the negative $x$-direction ($p_x<0$) are accounted for.}
\label{fig:scalelength-spectra}
\end{figure}

The proton spectra for simulations of four different sheet scale lengths are shown in Figure \ref{fig:scalelength-spectra} at three different times. The spectra show no significant difference for all but the thickest target. The spectrum corresponding to the simulation shown in Figure \ref{fig:scalelength} however shows a slight less pronounced peak towards the end of the simulation, compared to the simulations of the thinner targets. Finally, we make note of the completely different shape of the spectra for the thickest target ($\lambda_0/2$), where from the start no proton acceleration can be seen, other than that due to the thermal expansion of the plasma.

\begin{figure}[!t]
  \includegraphics[width=\columnwidth]{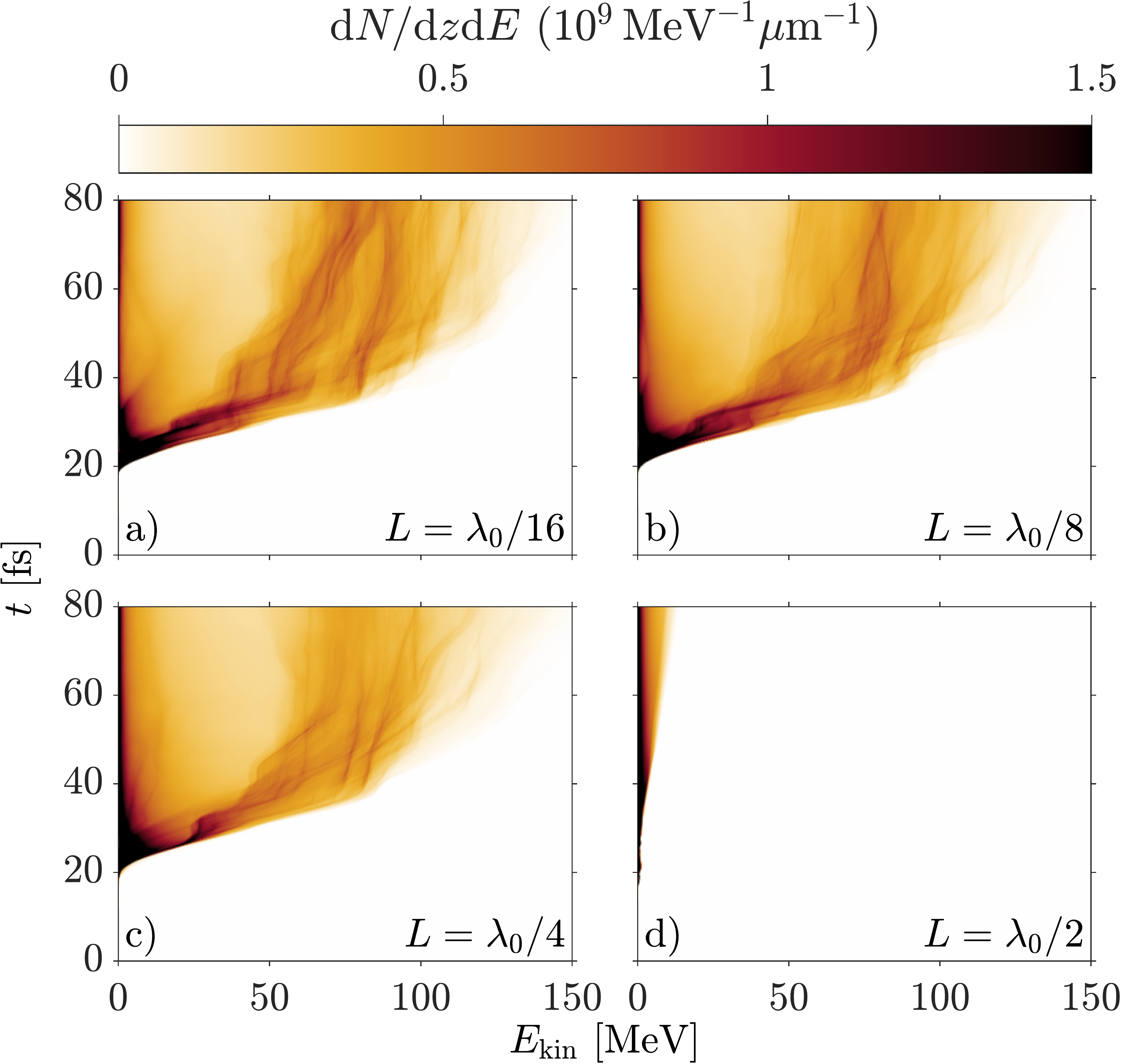}
\caption{Time evolution of the spectra of protons from the thin plasma layer for different sheet scale lengths, $L$. Only protons within a distance of half a pulse waist from the $x$-axis ($|y| < w/2$) and moving in the negative $x$-direction ($p_x<0$) are accounted for.}
\label{fig:scalelength-tspectra}
\end{figure}

The time evolution of the proton spectra are presented in Figure \ref{fig:scalelength-tspectra}, showing their evolution in much greater detail. We can clearly see that the spectral evolution is essentially unaffected by changes in the sheet scale length for $L \leq \lambda_0/4$, Figures \ref{fig:scalelength-tspectra}(a,b,c), as the only visible difference between these spectra is that the peak gets slightly less pronounced for thicker targets. We also see that that the spectral evolution corresponding to the thickest target, Figure \ref{fig:scalelength-tspectra}(d), is completely different and shows only a thermal evolution at low energies, similar to what would be expected of a heated plasma. 

The rapid transition of the outcome for scale lengths exceeding $\lambda_0/4$, together with its insensitivity for scale lengths on the order of and below this value, indicate that there is a threshold effect in which CSWA is enabled when the target is thinner than some limiting value. For scale lengths on the order of the width of the ponderomotive potential ($\sim \lambda_0/2$) the electrons will be insufficiently locked to the field node. Because of the dramatically different electron dynamics most of them are lost, thereby impeding the acceleration of the protons. However, in the regime where $L \lesssim \lambda_0$ we expect the CSWA scheme to continue to perform efficiently independently of the scale length, as it is the areal density that determines the threshold for relativistic transparency and this remains largely unaffected by the thermal expansion. Unless the pre-transparency interaction moves the electrons such that they get further than about $\lambda_0/4$ from the field node, they will be compressed and locked to the node. The proton acceleration will then commence as previously described for an infinitely thin sheet.

\section{Oblique incidence}\label{sec:angle}

Experiments on the interaction of intense lasers with solid targets are frequently restricted to oblique incidence, in order to protect the laser system from backreflections. The tolerance of CSWA to oblique incidence angles can therefore be of importance. For this reason we here perform a study similar to that in Section \ref{sec:thickness}, but now with respect to the incidence angle of the laser pulse.

The incident field is thus rotated about the $z$-axis by an angle $\theta$ such that the phase is now given by $\eta = t-(x\cos{\theta}+y\sin{\theta})/c$ and
\begin{equation}
\begin{aligned}
\vec{E} &= \frac{a_0}{\sqrt{2}}\Psi(\eta) [-\sin{\theta}\hat x + \cos{\theta}\hat y + i\hat z ], \\
\vec{B} &= (\cos{\theta}\hat{x}+\sin{\theta}\hat{y}) \times \vec{E}.
\end{aligned}
\end{equation}

\begin{figure}[!t]
\centering
  \includegraphics[width=\columnwidth]{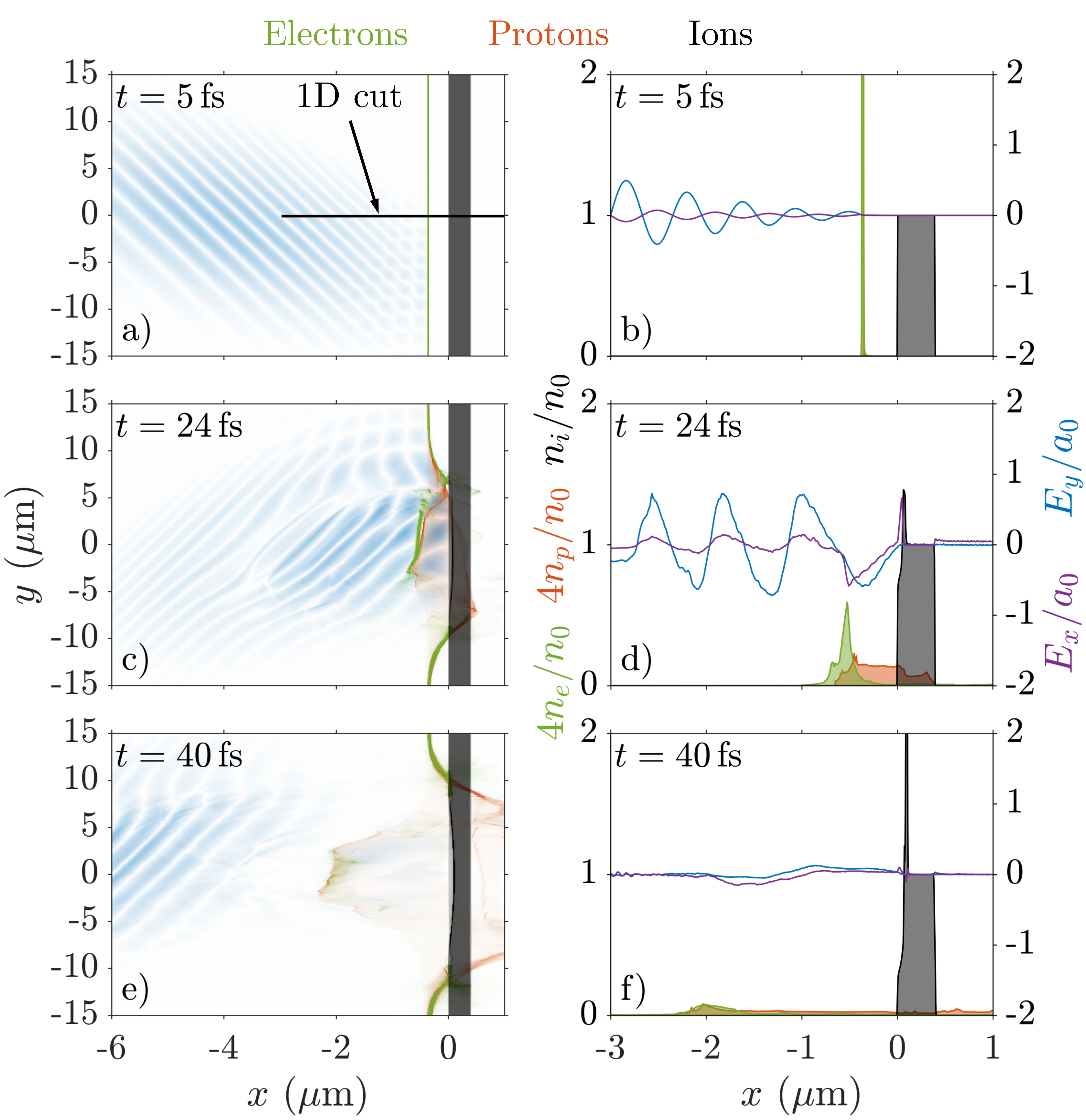}
\caption{A 2D PIC simulation for a laser energy $\varepsilon_0 = \unit[80]{J}$, bandwidth $\Delta\omega = 0.5\omega_0$ and chirp $\mathcal{C} = -3.5$ . The laser pulse is incident on the target at an angle of $\theta = 10^\circ$. Note also that the two coordinate axes of (a,c,e) are scaled differently (the $x$-axis is stretched about $6\times$) making angles to the vertical exaggerated. For this reason the wavefronts appear tilted compared to the propagation axis, but are in fact not. The information is presented as in Figure \ref{fig:ideal}.}
\label{fig:angle}
\end{figure}

In Figure \ref{fig:angle} the case of an incidence angle of $10^\circ$ is shown for several time instances both in the simulated 2D space and in a 1D cut along the $x$-axis. The simulations were performed for a circularly polarized pulse with laser energy, bandwidth and spot size as before ($\varepsilon_0 = \unit[80]{J}$, $\Delta\omega = 0.5\omega_0$, $w = \unit[10]{{\mu}m}$) but now with chirp $\mathcal{C} = -3.5$. The areal density of the thin sheet is $5.4\sigma_\mathrm{cr}$.

Figure \ref{fig:angle}(a-b) shows the simulation some time before the relativistic transparency of the thin sheet sets in and the $x$-component of the electric field can now be seen in Figure \ref{fig:angle}(b). In Figure \ref{fig:angle}(c-d) the electrons within the laser spot ($|y|<w/2$) are again seen to be locked at the field node but that the sheet is now tilted. For this reason, the proton and electron densities appear less peaked in Figure \ref{fig:angle}(d) compared to Figure \ref{fig:scalelength}(d) and Figure \ref{fig:ideal}(d), as they are averaged over the spot size. However, despite the oblique incidence, the electrostatic field, $E_x$, looks much like for the case of normal incidence, along $y = 0$, disregarding the more rapidly varying contribution of the laser field. Furthermore, in Figure \ref{fig:angle}(e-f) the protons are seen to be accelerated via a residual field after the laser pulse has passed, similarly to the ideal case of Figure \ref{fig:ideal}(e-f).

\begin{figure}[!t]
  \includegraphics[width=\columnwidth]{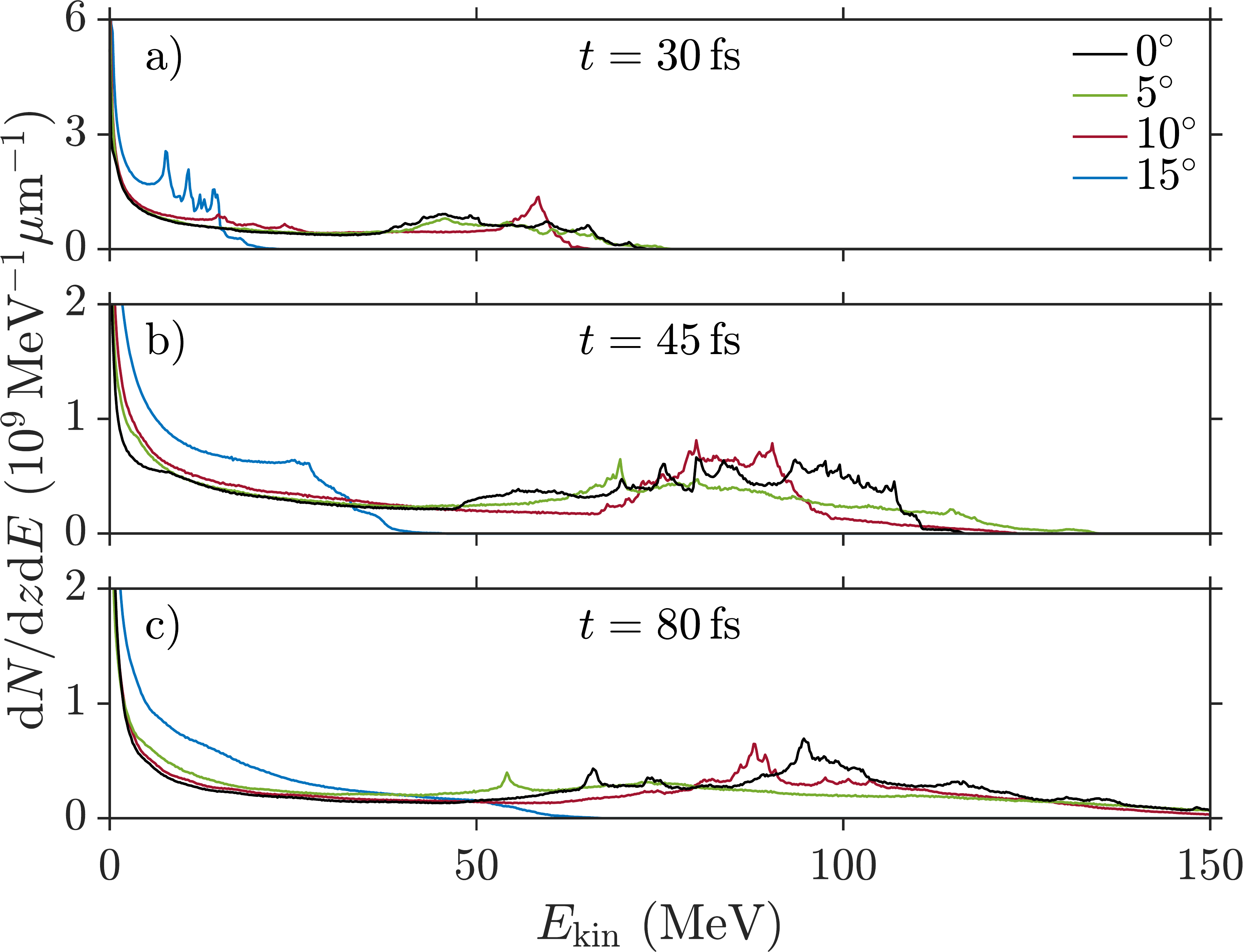}
\caption{Spectra of protons from the thin plasma layer for different incidence angles, $\theta$,  shown at three times $\unit[30]{fs}$ (a), $\unit[45]{fs}$ (b) and $\unit[80]{fs}$ (c). Only protons within a distance of half a pulse waist from the $x$-axis ($|y| < w/2$) and moving in the negative $x$-direction ($p_x<0$) are accounted for.}
\label{fig:angle-spectra}
\end{figure}

The corresponding proton spectra are shown in Figure \ref{fig:angle-spectra} for four different incidence angles and at three different instances in time. The spectra show little difference between incidence angles of $0^\circ$, $5^\circ$ and $10^\circ$ at early times except for the $10^\circ$ being more strongly peaked at the end of the tail. Their spectra are then seen to evolve similarly in time, toward higher energies and ultimately broadening. The spectrum for the $5^\circ$ case can however be seen to deviate somewhat from the other two at later times, becoming broader and less strongly peaked than both the $0^\circ$ and $10^\circ$ spectra. This was found to be due to a filamentation of the accelerated sheet occuring after the electrons had been released. Furthermore, the spectrum from the $15^\circ$ incidence angle simulation is seen to be drastically different from the ones with more moderate incidence angles, initially displaying several sharp peaks, but at much lower energies than the other spectra and due to a TNSA-like acceleration of the leading protons the high-end tail of this spectrum tends to a more thermal shape. 

\begin{figure}[!t]
  \includegraphics[width=\columnwidth]{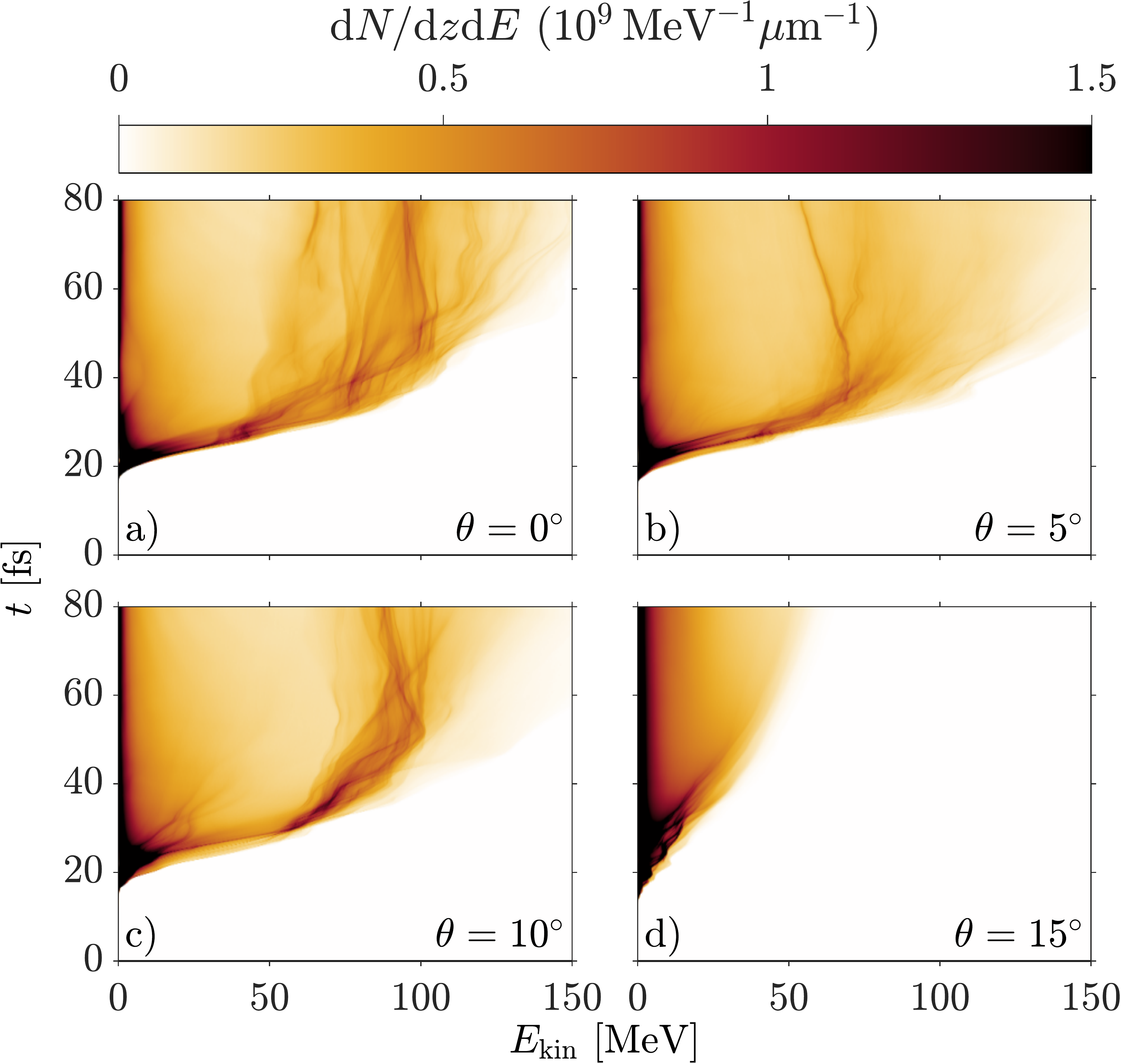}
\caption{Time evolution of the spectra of protons from the thin plasma layer for different incidence angles, $\theta$. Only protons within a distance of half a pulse waist from the $x$-axis ($|y| < w/2$) and moving in the negative $x$-direction ($p_x<0$) are accounted for.}
\label{fig:angle-tspectra}
\end{figure}

Looking at a more detailed time evolution of the proton spectra, Figure \ref{fig:angle-tspectra}, we can more clearly see the differences, but also the similarities, between the $5^\circ$ and the $0^\circ$ and $10^\circ$ cases. It further shows that their spectral evolution is initially very much the same before the electrons are released at around $\unit[40]{fs}$, after which point the spectra start to spread out. While the $5^\circ$ spectrum then displays a slightly less pronounced peak and a slightly lower total particle number, its general shape remains similar to the other two. For an incidence angle of $15^\circ$ however the spectrum instead shows a TNSA-like evolution after some initial acceleration period.

A maximum acceptable angle of incidence can be expressed in terms of the smallest $f$-number allowed, such that backreflections onto the focusing optics can be avoided. For an ideal Gaussian pulse at $5^\circ$ this would correspond to a minimum $f$-number of $f/5.2$. Since the acceleration scheme does not rely on tight focusing, but rather requires the spot size to be much greater than the laser wavelength, the use of large $f$-numbers will not be a limiting factor as long as a sufficient laser amplitude can still be accessed.

\section{Elliptical Polarization}\label{sec:pol}
In making the laser field circularly polarized a quarter-waveplate is often used. For laser pulses of very large bandwidth, which is preferable for efficient use of the CSWA scheme, this may present certain difficulties as the bandwidth over which ordinary waveplates can provide roughly the correct relative phase change is limited. Furthermore, the acceleration scheme rests on the fact that the electrons are locked in the longitudinal direction by the ponderomotive force of the standing wave and in the transverse directions due to the circular polarization of the laser field. In the case of a linearly polarized laser field the locking in the transverse direction will fail. Under such circumstances the electron layer will instead be heated and the insufficient trapping in the transverse direction will be unable to keep instabilities from forming, ultimately ruining the proton acceleration.

We here study an elliptically polarized laser pulse propagating in the longitudinal ($x$) direction chirped linearly according to equation \eqref{eq:ChirpedPulse}
\begin{equation}
\vec{E} = a_0\Psi(\eta) [\cos(\mathcal{E}) \hat y + i\sin(\mathcal{E}) \hat z ] ,\quad 
\vec{B} = \hat{x} \times \vec{E},
\end{equation}
where an ellipticity angle $\mathcal{E}$ of $0^\circ$, $45^\circ$ and $90^\circ$ corresponds to linear $y$, circular and linear $z$ polarization, respectively. The ellipticity angle is related to the ellipticity $\varepsilon$, defined as the axial ratio, as $\tan\mathcal{E} = \varepsilon$. As we are interested only in how sensitive the scheme is to different ellipticity angles in the viscinity of circular polarization, we here focus on ellipticity angles centered around $45^\circ$.

We again performed 2D simulations, now varying the ellipticity angle $\mathcal{E}$ in order to ascertain the limit to the CSWA scheme of the ellipticity. The simulations were performed for a laser energy, bandwidth, spot size and chirp as in section \ref{sec:thickness} ($\varepsilon_0 = \unit[80]{J}$, $\Delta\omega = 0.5\omega_0$, $w = \unit[10]{{\mu}m}$, $\mathcal{C} = -4$). Similarly, the areal density of the thin sheet is $5\sigma_\mathrm{cr}$.

\begin{figure}[!t]
  \includegraphics[width=\columnwidth]{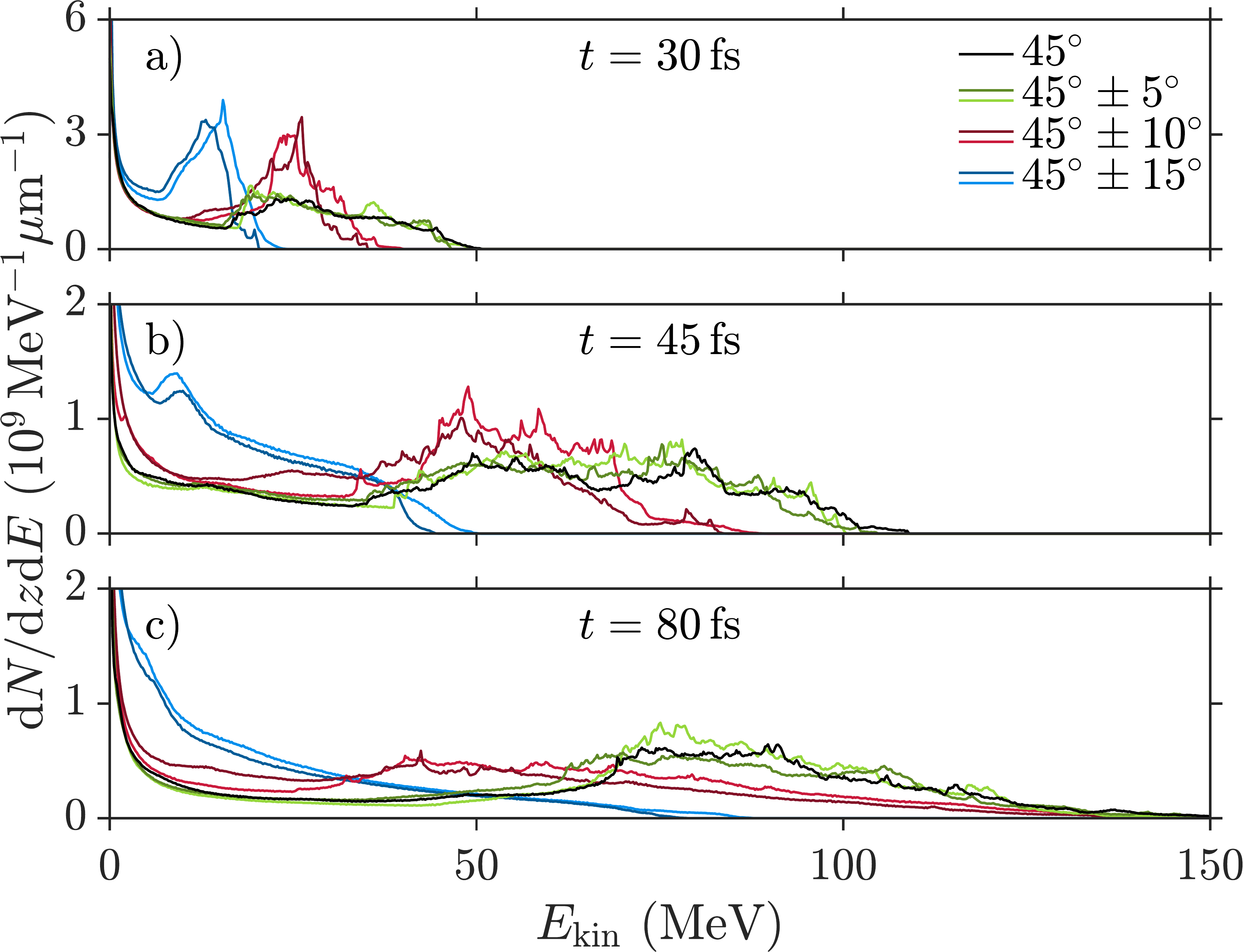}
\caption{Spectra of protons from the thin plasma layer for different ellipticity angles, $\mathcal{E}$,  shown at three times $\unit[30]{fs}$ (a), $\unit[45]{fs}$ (b) and $\unit[80]{fs}$ (c). Only protons within a distance of half a pulse waist from the laser propagation axis ($|y| < w/2$) and moving in the negative $x$-direction ($p_x<0$) are accounted for.}
\label{fig:ellipticity-spectra}
\end{figure}

The proton spectra of the simulations are presented in Figure \ref{fig:ellipticity-spectra} for seven different ellipticity angles and at three different times. The spectra show that there is no qualitative difference between ellipticities where the major polarization axis is in or out of the simulation plane. Furthermore, it can be seen that the spectra for large ellipticity angles are initially much more strongly peaked than for more moderate angles. The $\mathcal{E}=45^\circ\pm10^\circ$ spectra retains their peak and also gets shifted to higher energies between $30$\,-\,$\unit[45]{fs}$ during the post acceleration through the residual field, but its leading edge is then substantially stretched out. The $\mathcal{E}=45^\circ\pm15^\circ$ spectra however evolves into a more thermal spectra as time progresses and completely losing their initially peaked shape, similar to what we saw in section \ref{sec:angle} for incidence angles deviating too much from the target normal. Finally, we note that the spectra for moderate ellipticity angles ($\mathcal{E}=45^\circ\pm5^\circ$) does not in any significant way differ from the case of perfect circular polarization $\mathcal{E}=45^\circ$.

\begin{figure}[!t]
  \includegraphics[width=\columnwidth]{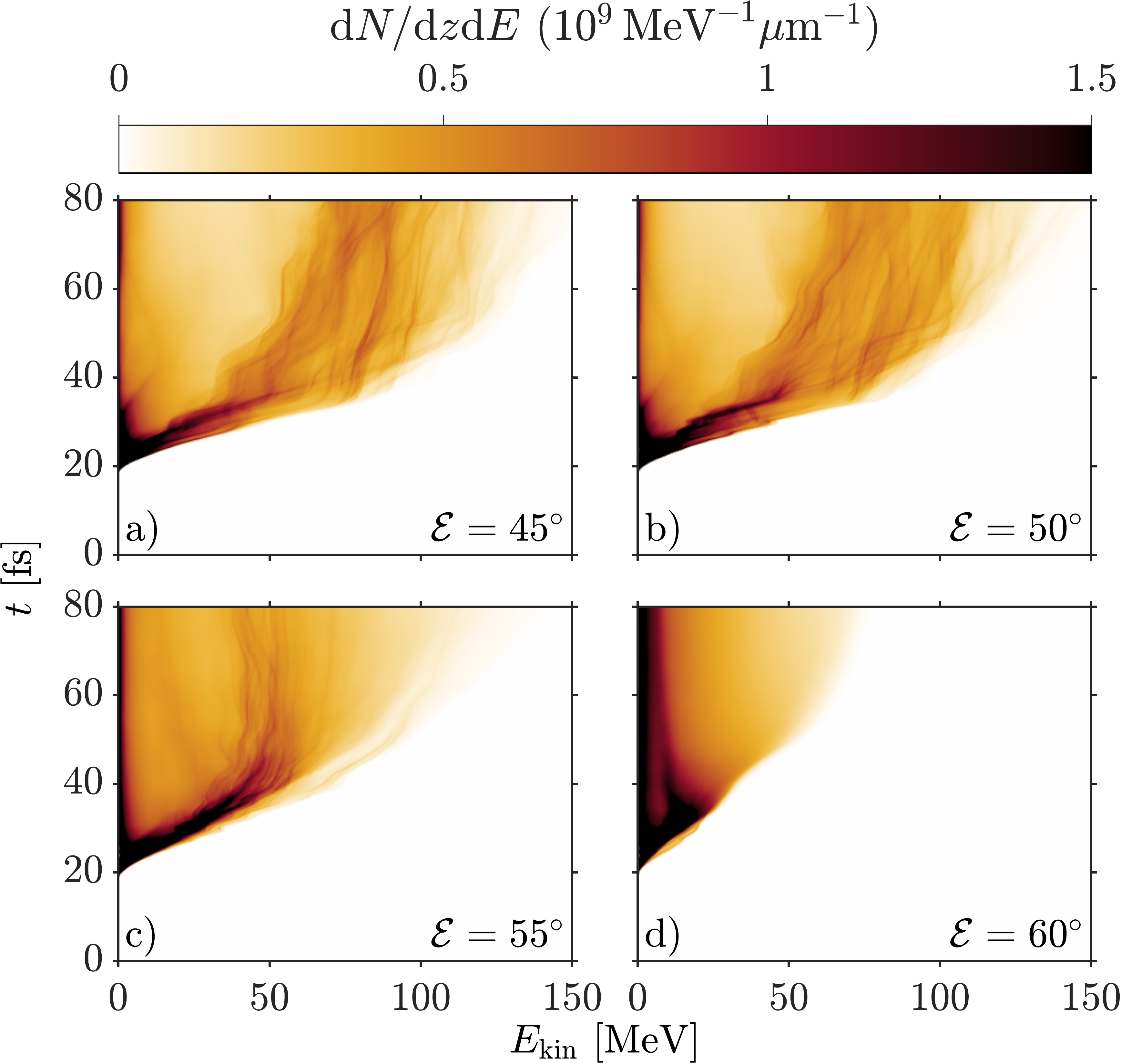}
\caption{Time evolution of the spectra of protons from the thin plasma layer for different ellipticity angles, $\mathcal{E}$. Only protons within a distance of half a pulse waist from the laser propagation axis ($|y| < w/2$) and moving in the negative $x$-direction ($p_x<0$) are accounted for.}
\label{fig:ellipticity-tspectra}
\end{figure}

In Figure \ref{fig:ellipticity-tspectra} we further show the time evolution of the proton spectra for $\mathcal{E} = 45^\circ, 50^\circ, 55^\circ, 60^\circ$. As noted earlier from Figure \ref{fig:ellipticity-spectra}, $\mathcal{E}=45^\circ$ and $\mathcal{E}=50^\circ$ are virtually indistinguishable while the peak for $\mathcal{E}=55^\circ$ becomes less pronounced. It also clearly shows a TNSA-like evolution of the $\mathcal{E}=60^\circ$ spectrum after around $\unit[50]{fs}$ as well as for the high energy tail of the $\mathcal{E}=55^\circ$ spectrum.

An ellipticity angle of $\mathcal{E} = 45^\circ\pm5^\circ$ corresponds to an ellipticity of $\varepsilon = 1.19$ (major over minor axis). This is well within current experimental limits and should be possible to achieve also for laser pulses of very large bandwidths, relying on for example achromatic waveplates.

\section{Prospects and limitations of CSWA}
The CSWA scheme is based on the frequency variation of the incoming laser field. The pulse bandwidth is therefore of great importance as it, together with the laser central frequency, uniquely determines the frequencies present in the laser radiation. The ratio of these two parameters, the fractional bandwidth, can be related to the optimally compressed pulse duration $\tau_0$ according to equation \eqref{eq:duration} and is a parameter that is central for the efficiency of the scheme.

For the acceleration of the ions to be efficient we must have that the ponderomotive locking of the electrons must be able to overcome the electrostatic charge separation force between the electrons and the ion sheet. We formulate this criterion by requiring that the electron deviation from the node position $\Delta x$ must be smaller than the maximum node displacement. This deviation,
\begin{equation}
\Delta x = \frac{\sqrt{2}\pi c (1+\mathcal{C}^2)^{1/4}}{a_0 \omega_0},
\end{equation}
was derived in Ref.\,\cite{Mackenroth} by equating the ponderomotive and Coulomb forces acting on an electron in the standing wave. The maximal displacement of the node is restricted by the spectral components of the pulse. More quantitatively it is restricted to half the wavelength difference, $\Delta\lambda$, between the largest and smallest wavelengths supported by the pulse. For a pulse of central frequency $\omega_0$ and bandwidth $\Delta\omega$ this wavelength difference is
\begin{equation}
\Delta\lambda = \frac{2\pi c}{\omega_0-\Delta\omega/2} - \frac{2\pi c}{\omega_0+\Delta\omega/2},
\end{equation}
which for $\Delta\omega/\omega_0 \ll 1$ to leading order becomes
\begin{equation}
\Delta\lambda = \lambda_0 \frac{\Delta\omega}{\omega_0}.
\end{equation}

We are also restricted by the fact that the mirror must be able to reflect the laser pulse as well as withstand the radiation pressure. Even if the ions remain stationary the effective reflection point may be shifted due to the radiation pressure excerted on the electrons. Quantitatively we similarly require the maximum displacement of the reflection point, $\Delta l$, to be smaller than the maximal movement of the node. These two constraints
\begin{equation}\label{eq:conditions}
\Delta x < \Delta \lambda/2, \quad
\Delta l < \Delta\lambda/2,
\end{equation}
give us two conditions on the laser amplitude for when the acceleration process can be expected to perform efficiently.

Assuming that the light pressure excerted on the electrons is balanced by the Coulomb attraction to the ions, we obtain the displacement of the reflection point in terms of the laser intensity $I$ and electron density of the mirror $n_e$, 
\begin{equation}
\Delta l = \frac{1}{\sqrt{\pi}}\frac{\sqrt{I/c}}{n_e e},
\end{equation}
and the maximum displacement can then be evaluated using the laser peak intensity,
\begin{equation}
I_\mathrm{max} = \frac{c}{4\pi}\frac{a_0^2}{\sqrt{1+\mathcal{C}^2}}\left(\frac{m_ec\omega_0}{e}\right)^2.
\end{equation}
Finally, we obtain from the inequalities \eqref{eq:conditions}
\begin{equation}\label{eq:a_lim}
\sqrt{2}\left( \frac{\Delta\omega}{\omega_0} \right)^{-1} < \frac{a_0}{\sqrt[4]{1+\mathcal{C}^2}} < \frac{\pi}{2} \left( \frac{\Delta\omega}{\omega_0} \right) \left( \frac{n_e}{n_\mathrm{cr}} \right).
\end{equation}

The left inequality of equation \eqref{eq:a_lim} can be interpreted in terms of the minimum energy required in the laser pulse for a given fractional bandwidth. Similarly, the right inequality can be interpreted in terms of the minimum mirror density needed for a given laser energy and fractional bandwidth. 

Next, we estimate the maximum achievable energy of the protons from two different generic constraints: the duration of the acceleration process and the maximum distance the ions can travel during the acceleration. We here assume that for the whole acceleration process the protons are accelerated by a longitudinal electric field with a strength equal to the laser pulse peak amplitude, $a_0$. The simulations presented in \cite{Mackenroth} indicate that under optimal conditions, the field strength approaches this natural limit. We thus also assume that the thin sheet has an optimal areal density as described in \cite{Mackenroth}. In terms of the first constraint, we consider the protons to be accelerated by the electrostatic field for the duration of the optimal acceleration time, also described in \cite{Mackenroth}. We can then estimate the maximal gain of momentum and interpret this in terms of the maximal energy gained by the protons:
\begin{equation}
E_\mathrm{T} \approx 0.0035 a_0^4 \left( \frac{\Delta\omega}{\omega_0}\right )^{-2} \,\mathrm{[MeV]}.
\end{equation}
In terms of the second constraint, we consider the protons to instead be accelerated by the electrostatic field for a maximum distance of $n\Delta\lambda/2$, from which we obtain a different limit to the energy gained by the protons
\begin{equation}
E_\mathrm{D} \approx 6.4a_0 \left( \frac{n}{2}\frac{\Delta\omega}{\omega_0} \right) \, \mathrm{[MeV]},
\end{equation}
where $n$ is the trapping node number. The minimum of these two expressions then gives us an estimate for the maximum achievable proton energy. 

From this estimate, we can clearly see that the constraint related to the maximal distance of acceleration appears as the main restriction for the majority of currently available high-intensity laser systems, as the typical bandwidth is on the order of a couple of percent and with energies of a few Joules. In order to take full advantage of the CSWA scheme new laser systems focusing on very high bandwidth would be necessary. This means that the development of modifications to the CSWA that can mitigate or overcome this restriction has great potential and is of high demand. At the same time, CSWA appears as a basic implementation, which can be used for a proof-of-principle demonstration of the SWA approach. This would be an essential step towards implementation of more advanced schemes for controlling the position of locked electrons in the future. In order for this to be possible with either current or upcoming laser systems the effects of CSWA must be made distinguishable from competing acceleration mechanisms such as TNSA. As the locked electrons are released the standing-wave-accelerated ion bunch undergoes TNSA-like acceleration as it propagates through space. For low bandwidths this presents certain difficulties as this translates into low ion energies, making them susceptible to the sheath fields generated by the motion of the co-propagating electrons and the original features in the ion energy spectrum are smeared out as a result. This can partially be overcome by instead accelerating the ion bunch towards the mirror, which can be done by simply reversing the sign of the chirp. In this case, the ion bunch enters the plasma of the mirror where they become shielded from the TNSA-like accelerating fields.

\begin{figure}[!t]
  \includegraphics[width=\columnwidth]{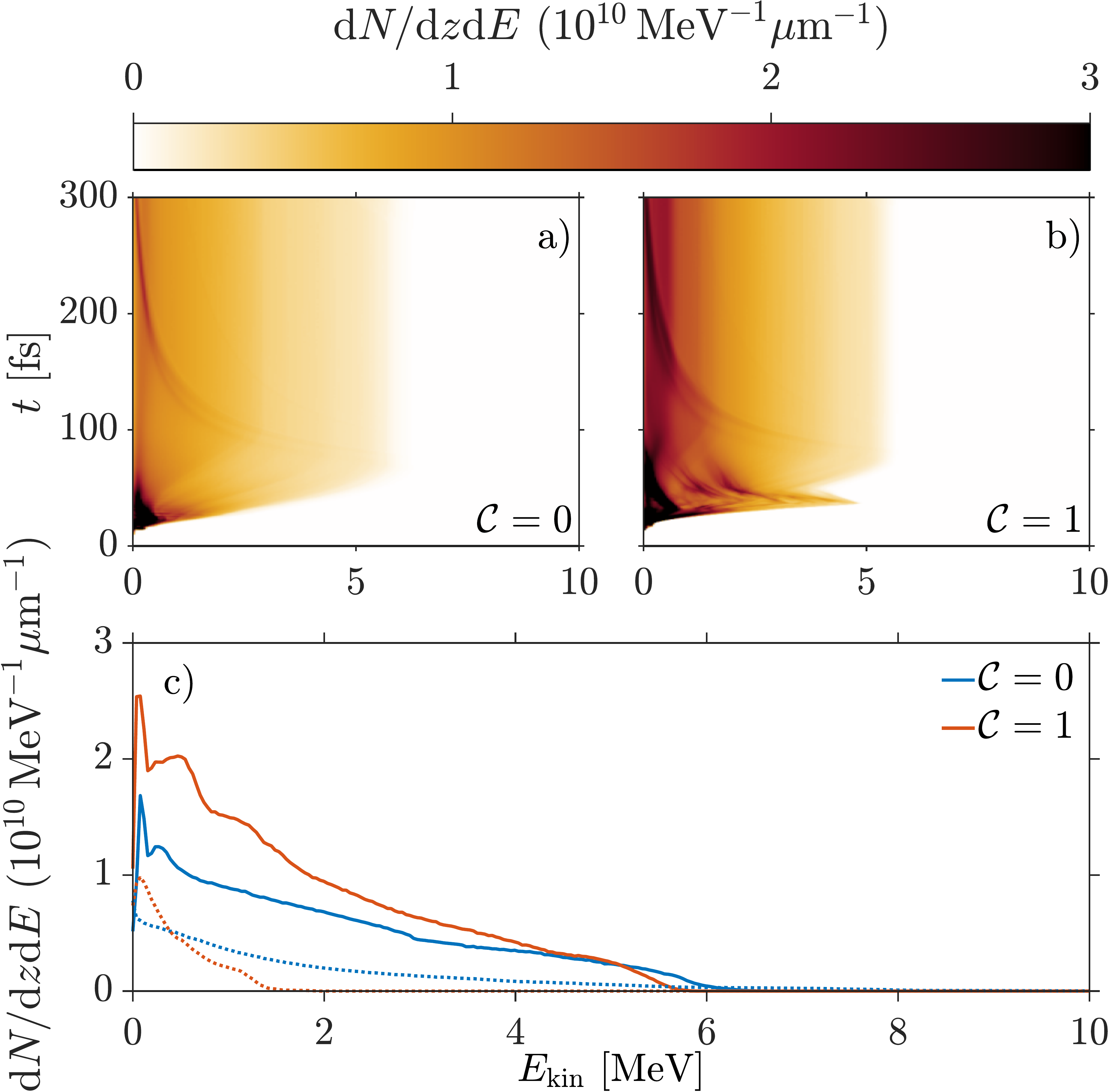}
\caption{Spectra of ions from the thin sheet for two different values of the chirp parameter obtained from 2D CSWA simulations with laser wavelength $\lambda_0 = \unit[810]{nm}$, energy $\varepsilon_0 = \unit[80]{J}$, and optimally compressed pulse duration $\tau_0 = \unit[10]{fs}$ ($\Delta\omega = 0.12\omega_0$) focused to a $\unit[10]{{\mu}m}$ spot size. The time evolution of the spectra are shown for (a) $\mathcal{C} = 0$ (b) and $\mathcal{C} = 1$, for the ions moving in the forward (positive $x$) direction. (c) The spectra of ions travelling in the forward (solid) and backward (dotted) directions are shown at $t = \unit[300]{fs}$ for the two values of the chirp parameter. Only ions with a propagation direction within $5^\circ$ of the target normal are accounted for.}
\label{fig:carbon-tspectra}
\end{figure}

In Figure \ref{fig:carbon-tspectra} we show the ion energy spectra obtained from two 2D simulations where the thin sheet consists of electrons, protons and ions of charge $+e$ and mass $12m_p$. The proton density was $10\%$ of the ion density. The mirror is modelled as before, but is made substantially thicker in order to not subject the ion bunch to strong sheath fields on the rear side of the mirror. The simulations were carried out with and without chirp ($\mathcal{C} = 0, 1$) using a laser of wavelength $\lambda_0 = \unit[810]{nm}$, energy $\varepsilon_0 = \unit[80]{J}$, and unchirped pulse duration $\tau_0 = \unit[10]{fs}$ ($\Delta\omega = 0.12\omega_0$) focused to a $\unit[10]{{\mu}m}$ spot size.

As discussed above, Figure \ref{fig:carbon-tspectra} shows the TNSA-like acceleration of the ions and the effect of adding a chirp to the laser pulse. The effect of the chirp is to accelerate a larger amount of ions to high energies, as compared to TNSA. This creates a larger spectral gap between forward and backward moving ions, as well as increases the amount of forward accelerated ions. With the ions propagating through the mirror this also results in a greater survival of non-thermal spectral features. This could possibly be used as a principal probe of CSWA in upcoming experiments.

\section{Conclusions}
In summary, we have assessed the prospects of the SWA approach for laser-driven ion acceleration and demonstrated that the process of locking and displacing electrons is sufficiently robust. Using PIC simulations we considered CSWA scheme as a particular implementation of the SWA approach and revealed the degree of tolerance to limited laser contrast, imperfect polarization and oblique incidence. Quantitatively we have shown that the CSWA scheme can allow an oblique incidence of up to $10^\circ$ to the target normal without significantly affecting the proton acceleration.  We further showed that the laser pulse can be allowed to be elliptical polarized with ellipticity angles within $10^\circ$ of circular polarization, corresponding to an ellipticity of $1.4$ (major over minor axis). Combined, this means that no extreme polarization control will be necessary and that damaging backreflections can be avoided without having to leverage quality of the beam. The effect of limited laser contrast was ascertained by varying the scale length of plasma expansion. For scale lengths smaller than about a quarter of the laser wavelength the results are not significantly affected.

We further investigated the prospects and limitations of the CSWA scheme. We provided estimates of the energy scaling for the accelerated protons with respect to the laser energy and bandwidth. We revealed the significance of a large bandwidth for this implementation of the SWA approach. The requirement on having a large bandwidth can however be relaxed for heavier ions, with a lower charge to mass ratio, as they will be less prone to catching up to the locked electron sheet and can thus be accelerated by a longer pulse before the energy gain saturates. Our analysis opens up for the development of other implementations of the SWA approach, in order to avoid or mitigate the restrictions imposed by the bandwidth. Finally, we identified a possible probe of SWA for a proof-of-principle experiment based on the CSWA implementation using current laser systems.

\begin{acknowledgments}
The authors would like to thank the \textsc{Picador} development team, and S. Bastrakov in particular, for their invaluable technical support and T.~G. Blackburn and C.-G. Wahlstr\"om for helpful discussions. We also acknowledge the financial support by the Knut and Alice Wallenberg Foundation through the grant ``Plasma based compact ion sources'' (PLIONA) and the Swedish Research Council (grant 2017-05148). The simulations were performed on resources provided by the Swedish National Infrastructure for Computing (SNIC) at HPC2N.
\end{acknowledgments}

\bibliography{references}

\begin{thebibliography}{57}%
\makeatletter
\providecommand \@ifxundefined [1]{%
 \@ifx{#1\undefined}
}%
\providecommand \@ifnum [1]{%
 \ifnum #1\expandafter \@firstoftwo
 \else \expandafter \@secondoftwo
 \fi
}%
\providecommand \@ifx [1]{%
 \ifx #1\expandafter \@firstoftwo
 \else \expandafter \@secondoftwo
 \fi
}%
\providecommand \natexlab [1]{#1}%
\providecommand \enquote  [1]{``#1''}%
\providecommand \bibnamefont  [1]{#1}%
\providecommand \bibfnamefont [1]{#1}%
\providecommand \citenamefont [1]{#1}%
\providecommand \href@noop [0]{\@secondoftwo}%
\providecommand \href [0]{\begingroup \@sanitize@url \@href}%
\providecommand \@href[1]{\@@startlink{#1}\@@href}%
\providecommand \@@href[1]{\endgroup#1\@@endlink}%
\providecommand \@sanitize@url [0]{\catcode `\\12\catcode `\$12\catcode
  `\&12\catcode `\#12\catcode `\^12\catcode `\_12\catcode `\%12\relax}%
\providecommand \@@startlink[1]{}%
\providecommand \@@endlink[0]{}%
\providecommand \url  [0]{\begingroup\@sanitize@url \@url }%
\providecommand \@url [1]{\endgroup\@href {#1}{\urlprefix }}%
\providecommand \urlprefix  [0]{URL }%
\providecommand \Eprint [0]{\href }%
\providecommand \doibase [0]{http://dx.doi.org/}%
\providecommand \selectlanguage [0]{\@gobble}%
\providecommand \bibinfo  [0]{\@secondoftwo}%
\providecommand \bibfield  [0]{\@secondoftwo}%
\providecommand \translation [1]{[#1]}%
\providecommand \BibitemOpen [0]{}%
\providecommand \bibitemStop [0]{}%
\providecommand \bibitemNoStop [0]{.\EOS\space}%
\providecommand \EOS [0]{\spacefactor3000\relax}%
\providecommand \BibitemShut  [1]{\csname bibitem#1\endcsname}%
\let\auto@bib@innerbib\@empty
\bibitem [{\citenamefont {Daido}\ \emph {et~al.}(2012)\citenamefont {Daido},
  \citenamefont {Nishiuchi},\ and\ \citenamefont {Pirozhkov}}]{Daido}%
  \BibitemOpen
  \bibfield  {author} {\bibinfo {author} {\bibfnamefont {H.}~\bibnamefont
  {Daido}}, \bibinfo {author} {\bibfnamefont {M.}~\bibnamefont {Nishiuchi}}, \
  and\ \bibinfo {author} {\bibfnamefont {A.~S.}\ \bibnamefont {Pirozhkov}},\
  }\href {\doibase 10.1088/0034-4885/75/5/056401} {\bibfield  {journal}
  {\bibinfo  {journal} {Rep. Prog. Phys.}\ }\textbf {\bibinfo {volume} {75}},\
  \bibinfo {pages} {056401} (\bibinfo {year} {2012})}\BibitemShut {NoStop}%
\bibitem [{\citenamefont {Macchi}\ \emph {et~al.}(2013)\citenamefont {Macchi},
  \citenamefont {Borghesi},\ and\ \citenamefont {Passoni}}]{Macchi}%
  \BibitemOpen
  \bibfield  {author} {\bibinfo {author} {\bibfnamefont {A.}~\bibnamefont
  {Macchi}}, \bibinfo {author} {\bibfnamefont {M.}~\bibnamefont {Borghesi}}, \
  and\ \bibinfo {author} {\bibfnamefont {M.}~\bibnamefont {Passoni}},\ }\href
  {\doibase 10.1103/RevModPhys.85.751} {\bibfield  {journal} {\bibinfo
  {journal} {Rev. Mod. Phys.}\ }\textbf {\bibinfo {volume} {85}},\ \bibinfo
  {pages} {751} (\bibinfo {year} {2013})}\BibitemShut {NoStop}%
\bibitem [{\citenamefont {Bulanov}\ \emph
  {et~al.}(2008{\natexlab{a}})\citenamefont {Bulanov}, \citenamefont {Brantov},
  \citenamefont {Bychenkov}, \citenamefont {Chvykov}, \citenamefont
  {Kalinchenko}, \citenamefont {Matsuoka}, \citenamefont {Rousseau},
  \citenamefont {Reed}, \citenamefont {Yanovsky}, \citenamefont {Krushelnick},
  \citenamefont {Litzenberg},\ and\ \citenamefont {Maksimchuk}}]{Bulanov}%
  \BibitemOpen
  \bibfield  {author} {\bibinfo {author} {\bibfnamefont {S.~S.}\ \bibnamefont
  {Bulanov}}, \bibinfo {author} {\bibfnamefont {A.}~\bibnamefont {Brantov}},
  \bibinfo {author} {\bibfnamefont {V.~Y.}\ \bibnamefont {Bychenkov}}, \bibinfo
  {author} {\bibfnamefont {V.}~\bibnamefont {Chvykov}}, \bibinfo {author}
  {\bibfnamefont {G.}~\bibnamefont {Kalinchenko}}, \bibinfo {author}
  {\bibfnamefont {T.}~\bibnamefont {Matsuoka}}, \bibinfo {author}
  {\bibfnamefont {P.}~\bibnamefont {Rousseau}}, \bibinfo {author}
  {\bibfnamefont {S.}~\bibnamefont {Reed}}, \bibinfo {author} {\bibfnamefont
  {V.}~\bibnamefont {Yanovsky}}, \bibinfo {author} {\bibfnamefont
  {K.}~\bibnamefont {Krushelnick}}, \bibinfo {author} {\bibfnamefont {D.~W.}\
  \bibnamefont {Litzenberg}}, \ and\ \bibinfo {author} {\bibfnamefont
  {A.}~\bibnamefont {Maksimchuk}},\ }\href {\doibase 10.1118/1.2900112}
  {\bibfield  {journal} {\bibinfo  {journal} {Med. Phys.}\ }\textbf {\bibinfo
  {volume} {35}},\ \bibinfo {pages} {1770} (\bibinfo {year}
  {2008}{\natexlab{a}})}\BibitemShut {NoStop}%
\bibitem [{\citenamefont {Hatchett}\ \emph {et~al.}(2000)\citenamefont
  {Hatchett}, \citenamefont {Brown}, \citenamefont {Cowan}, \citenamefont
  {Henry}, \citenamefont {Johnson}, \citenamefont {Key}, \citenamefont {Koch},
  \citenamefont {Langdon}, \citenamefont {Lasinski}, \citenamefont {Lee},
  \citenamefont {Mackinnon}, \citenamefont {Pennington}, \citenamefont {Perry},
  \citenamefont {Phillips}, \citenamefont {Roth}, \citenamefont {Sangster},
  \citenamefont {Singh}, \citenamefont {Snavely}, \citenamefont {Stoyer},
  \citenamefont {Wilks},\ and\ \citenamefont {Yasuike}}]{hatchett.pop.2000}%
  \BibitemOpen
  \bibfield  {author} {\bibinfo {author} {\bibfnamefont {S.~P.}\ \bibnamefont
  {Hatchett}}, \bibinfo {author} {\bibfnamefont {C.~G.}\ \bibnamefont {Brown}},
  \bibinfo {author} {\bibfnamefont {T.~E.}\ \bibnamefont {Cowan}}, \bibinfo
  {author} {\bibfnamefont {E.~A.}\ \bibnamefont {Henry}}, \bibinfo {author}
  {\bibfnamefont {J.~S.}\ \bibnamefont {Johnson}}, \bibinfo {author}
  {\bibfnamefont {M.~H.}\ \bibnamefont {Key}}, \bibinfo {author} {\bibfnamefont
  {J.~A.}\ \bibnamefont {Koch}}, \bibinfo {author} {\bibfnamefont {A.~B.}\
  \bibnamefont {Langdon}}, \bibinfo {author} {\bibfnamefont {B.~F.}\
  \bibnamefont {Lasinski}}, \bibinfo {author} {\bibfnamefont {R.~W.}\
  \bibnamefont {Lee}}, \bibinfo {author} {\bibfnamefont {A.~J.}\ \bibnamefont
  {Mackinnon}}, \bibinfo {author} {\bibfnamefont {D.~M.}\ \bibnamefont
  {Pennington}}, \bibinfo {author} {\bibfnamefont {M.~D.}\ \bibnamefont
  {Perry}}, \bibinfo {author} {\bibfnamefont {T.~W.}\ \bibnamefont {Phillips}},
  \bibinfo {author} {\bibfnamefont {M.}~\bibnamefont {Roth}}, \bibinfo {author}
  {\bibfnamefont {T.~C.}\ \bibnamefont {Sangster}}, \bibinfo {author}
  {\bibfnamefont {M.~S.}\ \bibnamefont {Singh}}, \bibinfo {author}
  {\bibfnamefont {R.~A.}\ \bibnamefont {Snavely}}, \bibinfo {author}
  {\bibfnamefont {M.~A.}\ \bibnamefont {Stoyer}}, \bibinfo {author}
  {\bibfnamefont {S.~C.}\ \bibnamefont {Wilks}}, \ and\ \bibinfo {author}
  {\bibfnamefont {K.}~\bibnamefont {Yasuike}},\ }\href {\doibase
  10.1063/1.874030} {\bibfield  {journal} {\bibinfo  {journal} {Physics of
  Plasmas}\ }\textbf {\bibinfo {volume} {7}},\ \bibinfo {pages} {2076}
  (\bibinfo {year} {2000})}\BibitemShut {NoStop}%
\bibitem [{\citenamefont {Wilks}\ \emph {et~al.}(2001)\citenamefont {Wilks},
  \citenamefont {Langdon}, \citenamefont {Cowan}, \citenamefont {Roth},
  \citenamefont {Singh}, \citenamefont {Hatchett}, \citenamefont {Key},
  \citenamefont {Pennington}, \citenamefont {MacKinnon},\ and\ \citenamefont
  {Snavely}}]{wilks.pop.2001}%
  \BibitemOpen
  \bibfield  {author} {\bibinfo {author} {\bibfnamefont {S.~C.}\ \bibnamefont
  {Wilks}}, \bibinfo {author} {\bibfnamefont {A.~B.}\ \bibnamefont {Langdon}},
  \bibinfo {author} {\bibfnamefont {T.~E.}\ \bibnamefont {Cowan}}, \bibinfo
  {author} {\bibfnamefont {M.}~\bibnamefont {Roth}}, \bibinfo {author}
  {\bibfnamefont {M.}~\bibnamefont {Singh}}, \bibinfo {author} {\bibfnamefont
  {S.}~\bibnamefont {Hatchett}}, \bibinfo {author} {\bibfnamefont {M.~H.}\
  \bibnamefont {Key}}, \bibinfo {author} {\bibfnamefont {D.}~\bibnamefont
  {Pennington}}, \bibinfo {author} {\bibfnamefont {A.}~\bibnamefont
  {MacKinnon}}, \ and\ \bibinfo {author} {\bibfnamefont {R.~A.}\ \bibnamefont
  {Snavely}},\ }\href {\doibase 10.1063/1.1333697} {\bibfield  {journal}
  {\bibinfo  {journal} {Physics of Plasmas}\ }\textbf {\bibinfo {volume} {8}},\
  \bibinfo {pages} {542} (\bibinfo {year} {2001})}\BibitemShut {NoStop}%
\bibitem [{\citenamefont {Mackinnon}\ \emph {et~al.}(2002)\citenamefont
  {Mackinnon}, \citenamefont {Sentoku}, \citenamefont {Patel}, \citenamefont
  {Price}, \citenamefont {Hatchett}, \citenamefont {Key}, \citenamefont
  {Andersen}, \citenamefont {Snavely},\ and\ \citenamefont
  {Freeman}}]{mackinnon.prl.2002}%
  \BibitemOpen
  \bibfield  {author} {\bibinfo {author} {\bibfnamefont {A.~J.}\ \bibnamefont
  {Mackinnon}}, \bibinfo {author} {\bibfnamefont {Y.}~\bibnamefont {Sentoku}},
  \bibinfo {author} {\bibfnamefont {P.~K.}\ \bibnamefont {Patel}}, \bibinfo
  {author} {\bibfnamefont {D.~W.}\ \bibnamefont {Price}}, \bibinfo {author}
  {\bibfnamefont {S.}~\bibnamefont {Hatchett}}, \bibinfo {author}
  {\bibfnamefont {M.~H.}\ \bibnamefont {Key}}, \bibinfo {author} {\bibfnamefont
  {C.}~\bibnamefont {Andersen}}, \bibinfo {author} {\bibfnamefont
  {R.}~\bibnamefont {Snavely}}, \ and\ \bibinfo {author} {\bibfnamefont
  {R.~R.}\ \bibnamefont {Freeman}},\ }\href {\doibase
  10.1103/PhysRevLett.88.215006} {\bibfield  {journal} {\bibinfo  {journal}
  {Phys. Rev. Lett.}\ }\textbf {\bibinfo {volume} {88}},\ \bibinfo {pages}
  {215006} (\bibinfo {year} {2002})}\BibitemShut {NoStop}%
\bibitem [{\citenamefont {Roth}\ \emph {et~al.}(2002)\citenamefont {Roth},
  \citenamefont {Blazevic}, \citenamefont {Geissel}, \citenamefont {Schlegel},
  \citenamefont {Cowan}, \citenamefont {Allen}, \citenamefont {Gauthier},
  \citenamefont {Audebert}, \citenamefont {Fuchs}, \citenamefont {Meyer-ter
  Vehn}, \citenamefont {Hegelich}, \citenamefont {Karsch},\ and\ \citenamefont
  {Pukhov}}]{Roth}%
  \BibitemOpen
  \bibfield  {author} {\bibinfo {author} {\bibfnamefont {M.}~\bibnamefont
  {Roth}}, \bibinfo {author} {\bibfnamefont {A.}~\bibnamefont {Blazevic}},
  \bibinfo {author} {\bibfnamefont {M.}~\bibnamefont {Geissel}}, \bibinfo
  {author} {\bibfnamefont {T.}~\bibnamefont {Schlegel}}, \bibinfo {author}
  {\bibfnamefont {T.~E.}\ \bibnamefont {Cowan}}, \bibinfo {author}
  {\bibfnamefont {M.}~\bibnamefont {Allen}}, \bibinfo {author} {\bibfnamefont
  {J.-C.}\ \bibnamefont {Gauthier}}, \bibinfo {author} {\bibfnamefont
  {P.}~\bibnamefont {Audebert}}, \bibinfo {author} {\bibfnamefont
  {J.}~\bibnamefont {Fuchs}}, \bibinfo {author} {\bibfnamefont
  {J.}~\bibnamefont {Meyer-ter Vehn}}, \bibinfo {author} {\bibfnamefont
  {M.}~\bibnamefont {Hegelich}}, \bibinfo {author} {\bibfnamefont
  {S.}~\bibnamefont {Karsch}}, \ and\ \bibinfo {author} {\bibfnamefont
  {A.}~\bibnamefont {Pukhov}},\ }\href {\doibase 10.1103/PhysRevSTAB.5.061301}
  {\bibfield  {journal} {\bibinfo  {journal} {Phys. Rev. ST Accel. Beams}\
  }\textbf {\bibinfo {volume} {5}},\ \bibinfo {pages} {061301} (\bibinfo {year}
  {2002})}\BibitemShut {NoStop}%
\bibitem [{\citenamefont {Mora}(2003)}]{Mora}%
  \BibitemOpen
  \bibfield  {author} {\bibinfo {author} {\bibfnamefont {P.}~\bibnamefont
  {Mora}},\ }\href {\doibase 10.1103/PhysRevLett.90.185002} {\bibfield
  {journal} {\bibinfo  {journal} {Phys. Rev. Lett.}\ }\textbf {\bibinfo
  {volume} {90}},\ \bibinfo {pages} {185002} (\bibinfo {year}
  {2003})}\BibitemShut {NoStop}%
\bibitem [{\citenamefont {Cowan}\ \emph {et~al.}(2004)\citenamefont {Cowan},
  \citenamefont {Fuchs}, \citenamefont {Ruhl}, \citenamefont {Kemp},
  \citenamefont {Audebert}, \citenamefont {Roth}, \citenamefont {Stephens},
  \citenamefont {Barton}, \citenamefont {Blazevic}, \citenamefont {Brambrink},
  \citenamefont {Cobble}, \citenamefont {Fern\'andez}, \citenamefont
  {Gauthier}, \citenamefont {Geissel}, \citenamefont {Hegelich}, \citenamefont
  {Kaae}, \citenamefont {Karsch}, \citenamefont {Le~Sage}, \citenamefont
  {Letzring}, \citenamefont {Manclossi}, \citenamefont {Meyroneinc},
  \citenamefont {Newkirk}, \citenamefont {P\'epin},\ and\ \citenamefont
  {Renard-LeGalloudec}}]{Cowan}%
  \BibitemOpen
  \bibfield  {author} {\bibinfo {author} {\bibfnamefont {T.~E.}\ \bibnamefont
  {Cowan}}, \bibinfo {author} {\bibfnamefont {J.}~\bibnamefont {Fuchs}},
  \bibinfo {author} {\bibfnamefont {H.}~\bibnamefont {Ruhl}}, \bibinfo {author}
  {\bibfnamefont {A.}~\bibnamefont {Kemp}}, \bibinfo {author} {\bibfnamefont
  {P.}~\bibnamefont {Audebert}}, \bibinfo {author} {\bibfnamefont
  {M.}~\bibnamefont {Roth}}, \bibinfo {author} {\bibfnamefont {R.}~\bibnamefont
  {Stephens}}, \bibinfo {author} {\bibfnamefont {I.}~\bibnamefont {Barton}},
  \bibinfo {author} {\bibfnamefont {A.}~\bibnamefont {Blazevic}}, \bibinfo
  {author} {\bibfnamefont {E.}~\bibnamefont {Brambrink}}, \bibinfo {author}
  {\bibfnamefont {J.}~\bibnamefont {Cobble}}, \bibinfo {author} {\bibfnamefont
  {J.}~\bibnamefont {Fern\'andez}}, \bibinfo {author} {\bibfnamefont {J.-C.}\
  \bibnamefont {Gauthier}}, \bibinfo {author} {\bibfnamefont {M.}~\bibnamefont
  {Geissel}}, \bibinfo {author} {\bibfnamefont {M.}~\bibnamefont {Hegelich}},
  \bibinfo {author} {\bibfnamefont {J.}~\bibnamefont {Kaae}}, \bibinfo {author}
  {\bibfnamefont {S.}~\bibnamefont {Karsch}}, \bibinfo {author} {\bibfnamefont
  {G.~P.}\ \bibnamefont {Le~Sage}}, \bibinfo {author} {\bibfnamefont
  {S.}~\bibnamefont {Letzring}}, \bibinfo {author} {\bibfnamefont
  {M.}~\bibnamefont {Manclossi}}, \bibinfo {author} {\bibfnamefont
  {S.}~\bibnamefont {Meyroneinc}}, \bibinfo {author} {\bibfnamefont
  {A.}~\bibnamefont {Newkirk}}, \bibinfo {author} {\bibfnamefont
  {H.}~\bibnamefont {P\'epin}}, \ and\ \bibinfo {author} {\bibfnamefont
  {N.}~\bibnamefont {Renard-LeGalloudec}},\ }\href {\doibase
  10.1103/PhysRevLett.92.204801} {\bibfield  {journal} {\bibinfo  {journal}
  {Phys. Rev. Lett.}\ }\textbf {\bibinfo {volume} {92}},\ \bibinfo {pages}
  {204801} (\bibinfo {year} {2004})}\BibitemShut {NoStop}%
\bibitem [{\citenamefont {Passoni}\ \emph {et~al.}(2010)\citenamefont
  {Passoni}, \citenamefont {Bertagna},\ and\ \citenamefont {Zani}}]{Passoni1}%
  \BibitemOpen
  \bibfield  {author} {\bibinfo {author} {\bibfnamefont {M.}~\bibnamefont
  {Passoni}}, \bibinfo {author} {\bibfnamefont {L.}~\bibnamefont {Bertagna}}, \
  and\ \bibinfo {author} {\bibfnamefont {A.}~\bibnamefont {Zani}},\ }\href
  {\doibase 10.1088/1367-2630/12/4/045012} {\bibfield  {journal} {\bibinfo
  {journal} {New J. Phys.}\ }\textbf {\bibinfo {volume} {12}},\ \bibinfo
  {pages} {045012} (\bibinfo {year} {2010})}\BibitemShut {NoStop}%
\bibitem [{\citenamefont {Ditmire}\ \emph {et~al.}(1997)\citenamefont
  {Ditmire}, \citenamefont {Tisch}, \citenamefont {Springate}, \citenamefont
  {Mason}, \citenamefont {Hay}, \citenamefont {Smith}, \citenamefont
  {Marangos},\ and\ \citenamefont {Hutchinson}}]{Ditmire}%
  \BibitemOpen
  \bibfield  {author} {\bibinfo {author} {\bibfnamefont {T.}~\bibnamefont
  {Ditmire}}, \bibinfo {author} {\bibfnamefont {J.}~\bibnamefont {Tisch}},
  \bibinfo {author} {\bibfnamefont {E.}~\bibnamefont {Springate}}, \bibinfo
  {author} {\bibfnamefont {M.}~\bibnamefont {Mason}}, \bibinfo {author}
  {\bibfnamefont {N.}~\bibnamefont {Hay}}, \bibinfo {author} {\bibfnamefont
  {R.}~\bibnamefont {Smith}}, \bibinfo {author} {\bibfnamefont
  {J.}~\bibnamefont {Marangos}}, \ and\ \bibinfo {author} {\bibfnamefont
  {M.}~\bibnamefont {Hutchinson}},\ }\href {\doibase 10.1038/386054a0}
  {\bibfield  {journal} {\bibinfo  {journal} {Nature}\ }\textbf {\bibinfo
  {volume} {386}},\ \bibinfo {pages} {54} (\bibinfo {year} {1997})}\BibitemShut
  {NoStop}%
\bibitem [{\citenamefont {Kovalev}\ \emph
  {et~al.}(2007{\natexlab{a}})\citenamefont {Kovalev}, \citenamefont
  {Bychenkov},\ and\ \citenamefont {Mima}}]{Kovalev1}%
  \BibitemOpen
  \bibfield  {author} {\bibinfo {author} {\bibfnamefont {V.~F.}\ \bibnamefont
  {Kovalev}}, \bibinfo {author} {\bibfnamefont {V.~Y.}\ \bibnamefont
  {Bychenkov}}, \ and\ \bibinfo {author} {\bibfnamefont {K.}~\bibnamefont
  {Mima}},\ }\href {\doibase 10.1063/1.2799160} {\bibfield  {journal} {\bibinfo
   {journal} {Phys. Plasmas}\ }\textbf {\bibinfo {volume} {14}},\ \bibinfo
  {pages} {103110} (\bibinfo {year} {2007}{\natexlab{a}})}\BibitemShut
  {NoStop}%
\bibitem [{\citenamefont {Kovalev}\ \emph
  {et~al.}(2007{\natexlab{b}})\citenamefont {Kovalev}, \citenamefont {Popov},
  \citenamefont {Bychenkov},\ and\ \citenamefont {Rozmus}}]{Kovalev2}%
  \BibitemOpen
  \bibfield  {author} {\bibinfo {author} {\bibfnamefont {V.~F.}\ \bibnamefont
  {Kovalev}}, \bibinfo {author} {\bibfnamefont {K.~I.}\ \bibnamefont {Popov}},
  \bibinfo {author} {\bibfnamefont {V.~Y.}\ \bibnamefont {Bychenkov}}, \ and\
  \bibinfo {author} {\bibfnamefont {W.}~\bibnamefont {Rozmus}},\ }\href
  {\doibase 10.1063/1.2731695} {\bibfield  {journal} {\bibinfo  {journal}
  {Phys. Plasmas}\ }\textbf {\bibinfo {volume} {14}},\ \bibinfo {pages}
  {053103} (\bibinfo {year} {2007}{\natexlab{b}})}\BibitemShut {NoStop}%
\bibitem [{\citenamefont {Esirkepov}\ \emph {et~al.}(2002)\citenamefont
  {Esirkepov}, \citenamefont {Bulanov}, \citenamefont {Nishihara},
  \citenamefont {Tajima}, \citenamefont {Pegoraro}, \citenamefont {Khoroshkov},
  \citenamefont {Mima}, \citenamefont {Daido}, \citenamefont {Kato},
  \citenamefont {Kitagawa}, \citenamefont {Nagai},\ and\ \citenamefont
  {Sakabe}}]{Esirkepov1}%
  \BibitemOpen
  \bibfield  {author} {\bibinfo {author} {\bibfnamefont {T.}~\bibnamefont
  {Esirkepov}}, \bibinfo {author} {\bibfnamefont {S.}~\bibnamefont {Bulanov}},
  \bibinfo {author} {\bibfnamefont {K.}~\bibnamefont {Nishihara}}, \bibinfo
  {author} {\bibfnamefont {T.}~\bibnamefont {Tajima}}, \bibinfo {author}
  {\bibfnamefont {F.}~\bibnamefont {Pegoraro}}, \bibinfo {author}
  {\bibfnamefont {V.}~\bibnamefont {Khoroshkov}}, \bibinfo {author}
  {\bibfnamefont {K.}~\bibnamefont {Mima}}, \bibinfo {author} {\bibfnamefont
  {H.}~\bibnamefont {Daido}}, \bibinfo {author} {\bibfnamefont
  {Y.}~\bibnamefont {Kato}}, \bibinfo {author} {\bibfnamefont {Y.}~\bibnamefont
  {Kitagawa}}, \bibinfo {author} {\bibfnamefont {K.}~\bibnamefont {Nagai}}, \
  and\ \bibinfo {author} {\bibfnamefont {S.}~\bibnamefont {Sakabe}},\ }\href
  {\doibase 10.1103/PhysRevLett.89.175003} {\bibfield  {journal} {\bibinfo
  {journal} {Phys. Rev. Lett.}\ }\textbf {\bibinfo {volume} {89}},\ \bibinfo
  {pages} {175003} (\bibinfo {year} {2002})}\BibitemShut {NoStop}%
\bibitem [{\citenamefont {Esirkepov}\ \emph {et~al.}(2006)\citenamefont
  {Esirkepov}, \citenamefont {Yamagiwa},\ and\ \citenamefont
  {Tajima}}]{Esirkepov2}%
  \BibitemOpen
  \bibfield  {author} {\bibinfo {author} {\bibfnamefont {T.}~\bibnamefont
  {Esirkepov}}, \bibinfo {author} {\bibfnamefont {M.}~\bibnamefont {Yamagiwa}},
  \ and\ \bibinfo {author} {\bibfnamefont {T.}~\bibnamefont {Tajima}},\ }\href
  {\doibase 10.1103/PhysRevLett.96.105001} {\bibfield  {journal} {\bibinfo
  {journal} {Phys. Rev. Lett.}\ }\textbf {\bibinfo {volume} {96}},\ \bibinfo
  {pages} {105001} (\bibinfo {year} {2006})}\BibitemShut {NoStop}%
\bibitem [{\citenamefont {Bulanov}\ \emph
  {et~al.}(2008{\natexlab{b}})\citenamefont {Bulanov}, \citenamefont {Brantov},
  \citenamefont {Bychenkov}, \citenamefont {Chvykov}, \citenamefont
  {Kalinchenko}, \citenamefont {Matsuoka}, \citenamefont {Rousseau},
  \citenamefont {Reed}, \citenamefont {Yanovsky}, \citenamefont {Litzenberg},
  \citenamefont {Krushelnick},\ and\ \citenamefont {Maksimchuk}}]{Bulanov2}%
  \BibitemOpen
  \bibfield  {author} {\bibinfo {author} {\bibfnamefont {S.~S.}\ \bibnamefont
  {Bulanov}}, \bibinfo {author} {\bibfnamefont {A.}~\bibnamefont {Brantov}},
  \bibinfo {author} {\bibfnamefont {V.~Y.}\ \bibnamefont {Bychenkov}}, \bibinfo
  {author} {\bibfnamefont {V.}~\bibnamefont {Chvykov}}, \bibinfo {author}
  {\bibfnamefont {G.}~\bibnamefont {Kalinchenko}}, \bibinfo {author}
  {\bibfnamefont {T.}~\bibnamefont {Matsuoka}}, \bibinfo {author}
  {\bibfnamefont {P.}~\bibnamefont {Rousseau}}, \bibinfo {author}
  {\bibfnamefont {S.}~\bibnamefont {Reed}}, \bibinfo {author} {\bibfnamefont
  {V.}~\bibnamefont {Yanovsky}}, \bibinfo {author} {\bibfnamefont {D.~W.}\
  \bibnamefont {Litzenberg}}, \bibinfo {author} {\bibfnamefont
  {K.}~\bibnamefont {Krushelnick}}, \ and\ \bibinfo {author} {\bibfnamefont
  {A.}~\bibnamefont {Maksimchuk}},\ }\href {\doibase
  10.1103/PhysRevE.78.026412} {\bibfield  {journal} {\bibinfo  {journal} {Phys.
  Rev. E}\ }\textbf {\bibinfo {volume} {78}},\ \bibinfo {pages} {026412}
  (\bibinfo {year} {2008}{\natexlab{b}})}\BibitemShut {NoStop}%
\bibitem [{\citenamefont {Yin}\ \emph {et~al.}(2006)\citenamefont {Yin},
  \citenamefont {Albright}, \citenamefont {Hegelich},\ and\ \citenamefont
  {Fern{\'a}ndez}}]{yin.lpb.2006}%
  \BibitemOpen
  \bibfield  {author} {\bibinfo {author} {\bibfnamefont {L.}~\bibnamefont
  {Yin}}, \bibinfo {author} {\bibfnamefont {B.~J.}\ \bibnamefont {Albright}},
  \bibinfo {author} {\bibfnamefont {B.~M.}\ \bibnamefont {Hegelich}}, \ and\
  \bibinfo {author} {\bibfnamefont {J.~C.}\ \bibnamefont {Fern{\'a}ndez}},\
  }\href {https://doi.org/10.1017/S0263034606060459} {\bibfield  {journal}
  {\bibinfo  {journal} {Laser and Particle Beams}\ }\textbf {\bibinfo {volume}
  {24}},\ \bibinfo {pages} {291} (\bibinfo {year} {2006})}\BibitemShut
  {NoStop}%
\bibitem [{\citenamefont {Yin}\ \emph {et~al.}(2007)\citenamefont {Yin},
  \citenamefont {Albright}, \citenamefont {Hegelich}, \citenamefont {Bowers},
  \citenamefont {Flippo}, \citenamefont {Kwan},\ and\ \citenamefont
  {Fern{\'a}ndez}}]{yin.pop.2007}%
  \BibitemOpen
  \bibfield  {author} {\bibinfo {author} {\bibfnamefont {L.}~\bibnamefont
  {Yin}}, \bibinfo {author} {\bibfnamefont {B.~J.}\ \bibnamefont {Albright}},
  \bibinfo {author} {\bibfnamefont {B.~M.}\ \bibnamefont {Hegelich}}, \bibinfo
  {author} {\bibfnamefont {K.~J.}\ \bibnamefont {Bowers}}, \bibinfo {author}
  {\bibfnamefont {K.~A.}\ \bibnamefont {Flippo}}, \bibinfo {author}
  {\bibfnamefont {T.~J.~T.}\ \bibnamefont {Kwan}}, \ and\ \bibinfo {author}
  {\bibfnamefont {J.~C.}\ \bibnamefont {Fern{\'a}ndez}},\ }\href {\doibase
  10.1063/1.2436857} {\bibfield  {journal} {\bibinfo  {journal} {Physics of
  Plasmas}\ }\textbf {\bibinfo {volume} {14}},\ \bibinfo {pages} {056706}
  (\bibinfo {year} {2007})}\BibitemShut {NoStop}%
\bibitem [{\citenamefont {Schlegel}\ \emph {et~al.}(2009)\citenamefont
  {Schlegel}, \citenamefont {Naumova}, \citenamefont {Tikhonchuk},
  \citenamefont {Labaune}, \citenamefont {Sokolov},\ and\ \citenamefont
  {Mourou}}]{Schlegel}%
  \BibitemOpen
  \bibfield  {author} {\bibinfo {author} {\bibfnamefont {T.}~\bibnamefont
  {Schlegel}}, \bibinfo {author} {\bibfnamefont {N.}~\bibnamefont {Naumova}},
  \bibinfo {author} {\bibfnamefont {V.~T.}\ \bibnamefont {Tikhonchuk}},
  \bibinfo {author} {\bibfnamefont {C.}~\bibnamefont {Labaune}}, \bibinfo
  {author} {\bibfnamefont {I.~V.}\ \bibnamefont {Sokolov}}, \ and\ \bibinfo
  {author} {\bibfnamefont {G.}~\bibnamefont {Mourou}},\ }\href {\doibase
  10.1063/1.3196845} {\bibfield  {journal} {\bibinfo  {journal} {Phys.
  Plasmas}\ }\textbf {\bibinfo {volume} {16}},\ \bibinfo {pages} {083103}
  (\bibinfo {year} {2009})}\BibitemShut {NoStop}%
\bibitem [{\citenamefont {Silva}\ \emph {et~al.}(2004)\citenamefont {Silva},
  \citenamefont {Marti}, \citenamefont {Davies}, \citenamefont {Fonseca},
  \citenamefont {Ren}, \citenamefont {Tsung},\ and\ \citenamefont
  {Mori}}]{Silva}%
  \BibitemOpen
  \bibfield  {author} {\bibinfo {author} {\bibfnamefont {L.~O.}\ \bibnamefont
  {Silva}}, \bibinfo {author} {\bibfnamefont {M.}~\bibnamefont {Marti}},
  \bibinfo {author} {\bibfnamefont {J.~R.}\ \bibnamefont {Davies}}, \bibinfo
  {author} {\bibfnamefont {R.~A.}\ \bibnamefont {Fonseca}}, \bibinfo {author}
  {\bibfnamefont {C.}~\bibnamefont {Ren}}, \bibinfo {author} {\bibfnamefont
  {F.~S.}\ \bibnamefont {Tsung}}, \ and\ \bibinfo {author} {\bibfnamefont
  {W.~B.}\ \bibnamefont {Mori}},\ }\href {\doibase
  10.1103/PhysRevLett.92.015002} {\bibfield  {journal} {\bibinfo  {journal}
  {Phys. Rev. Lett.}\ }\textbf {\bibinfo {volume} {92}},\ \bibinfo {pages}
  {015002} (\bibinfo {year} {2004})}\BibitemShut {NoStop}%
\bibitem [{\citenamefont {Haberberger}\ \emph {et~al.}(2012)\citenamefont
  {Haberberger}, \citenamefont {Tochitsky}, \citenamefont {Fiuza},
  \citenamefont {Gong}, \citenamefont {Fonseca}, \citenamefont {Silva},
  \citenamefont {Mori},\ and\ \citenamefont {Joshi}}]{Haberberger}%
  \BibitemOpen
  \bibfield  {author} {\bibinfo {author} {\bibfnamefont {D.}~\bibnamefont
  {Haberberger}}, \bibinfo {author} {\bibfnamefont {S.}~\bibnamefont
  {Tochitsky}}, \bibinfo {author} {\bibfnamefont {F.}~\bibnamefont {Fiuza}},
  \bibinfo {author} {\bibfnamefont {C.}~\bibnamefont {Gong}}, \bibinfo {author}
  {\bibfnamefont {R.~A.}\ \bibnamefont {Fonseca}}, \bibinfo {author}
  {\bibfnamefont {L.~O.}\ \bibnamefont {Silva}}, \bibinfo {author}
  {\bibfnamefont {W.~B.}\ \bibnamefont {Mori}}, \ and\ \bibinfo {author}
  {\bibfnamefont {C.}~\bibnamefont {Joshi}},\ }\href
  {http://dx.doi.org/10.1038/nphys2130} {\bibfield  {journal} {\bibinfo
  {journal} {Nature Phys.}\ }\textbf {\bibinfo {volume} {8}},\ \bibinfo {pages}
  {95} (\bibinfo {year} {2012})}\BibitemShut {NoStop}%
\bibitem [{\citenamefont {Bulanov}\ \emph {et~al.}(2005)\citenamefont
  {Bulanov}, \citenamefont {Dylov}, \citenamefont {Esirkepov}, \citenamefont
  {Kamenets},\ and\ \citenamefont {Sokolov}}]{bulanov.ppr.2005}%
  \BibitemOpen
  \bibfield  {author} {\bibinfo {author} {\bibfnamefont {S.~V.}\ \bibnamefont
  {Bulanov}}, \bibinfo {author} {\bibfnamefont {D.~V.}\ \bibnamefont {Dylov}},
  \bibinfo {author} {\bibfnamefont {T.~Z.}\ \bibnamefont {Esirkepov}}, \bibinfo
  {author} {\bibfnamefont {F.~F.}\ \bibnamefont {Kamenets}}, \ and\ \bibinfo
  {author} {\bibfnamefont {D.~V.}\ \bibnamefont {Sokolov}},\ }\href {\doibase
  10.1134/1.1925787} {\bibfield  {journal} {\bibinfo  {journal} {Plasma Physics
  Reports}\ }\textbf {\bibinfo {volume} {31}},\ \bibinfo {pages} {369}
  (\bibinfo {year} {2005})}\BibitemShut {NoStop}%
\bibitem [{\citenamefont {Nakamura}\ \emph {et~al.}(2010)\citenamefont
  {Nakamura}, \citenamefont {Bulanov}, \citenamefont {Esirkepov},\ and\
  \citenamefont {Kando}}]{nakamura.prl.2010}%
  \BibitemOpen
  \bibfield  {author} {\bibinfo {author} {\bibfnamefont {T.}~\bibnamefont
  {Nakamura}}, \bibinfo {author} {\bibfnamefont {S.~V.}\ \bibnamefont
  {Bulanov}}, \bibinfo {author} {\bibfnamefont {T.~Z.}\ \bibnamefont
  {Esirkepov}}, \ and\ \bibinfo {author} {\bibfnamefont {M.}~\bibnamefont
  {Kando}},\ }\href {\doibase 10.1103/PhysRevLett.105.135002} {\bibfield
  {journal} {\bibinfo  {journal} {Phys. Rev. Lett.}\ }\textbf {\bibinfo
  {volume} {105}},\ \bibinfo {pages} {135002} (\bibinfo {year}
  {2010})}\BibitemShut {NoStop}%
\bibitem [{\citenamefont {Esirkepov}\ \emph {et~al.}(2004)\citenamefont
  {Esirkepov}, \citenamefont {Borghesi}, \citenamefont {Bulanov}, \citenamefont
  {Mourou},\ and\ \citenamefont {Tajima}}]{Esirkepov3}%
  \BibitemOpen
  \bibfield  {author} {\bibinfo {author} {\bibfnamefont {T.}~\bibnamefont
  {Esirkepov}}, \bibinfo {author} {\bibfnamefont {M.}~\bibnamefont {Borghesi}},
  \bibinfo {author} {\bibfnamefont {S.}~\bibnamefont {Bulanov}}, \bibinfo
  {author} {\bibfnamefont {G.}~\bibnamefont {Mourou}}, \ and\ \bibinfo {author}
  {\bibfnamefont {T.}~\bibnamefont {Tajima}},\ }\href {\doibase
  10.1103/PhysRevLett.92.175003} {\bibfield  {journal} {\bibinfo  {journal}
  {Phys. Rev. Lett.}\ }\textbf {\bibinfo {volume} {92}},\ \bibinfo {pages}
  {175003} (\bibinfo {year} {2004})}\BibitemShut {NoStop}%
\bibitem [{\citenamefont {Bulanov}\ \emph {et~al.}(2010)\citenamefont
  {Bulanov}, \citenamefont {Echkina}, \citenamefont {Esirkepov}, \citenamefont
  {Inovenkov}, \citenamefont {Kando}, \citenamefont {Pegoraro},\ and\
  \citenamefont {Korn}}]{Bulanov3}%
  \BibitemOpen
  \bibfield  {author} {\bibinfo {author} {\bibfnamefont {S.~V.}\ \bibnamefont
  {Bulanov}}, \bibinfo {author} {\bibfnamefont {E.~Y.}\ \bibnamefont
  {Echkina}}, \bibinfo {author} {\bibfnamefont {T.~Z.}\ \bibnamefont
  {Esirkepov}}, \bibinfo {author} {\bibfnamefont {I.~N.}\ \bibnamefont
  {Inovenkov}}, \bibinfo {author} {\bibfnamefont {M.}~\bibnamefont {Kando}},
  \bibinfo {author} {\bibfnamefont {F.}~\bibnamefont {Pegoraro}}, \ and\
  \bibinfo {author} {\bibfnamefont {G.}~\bibnamefont {Korn}},\ }\href {\doibase
  10.1103/PhysRevLett.104.135003} {\bibfield  {journal} {\bibinfo  {journal}
  {Phys. Rev. Lett.}\ }\textbf {\bibinfo {volume} {104}},\ \bibinfo {pages}
  {135003} (\bibinfo {year} {2010})}\BibitemShut {NoStop}%
\bibitem [{\citenamefont {Henig}\ \emph {et~al.}(2009)\citenamefont {Henig},
  \citenamefont {Steinke}, \citenamefont {Schn\"urer}, \citenamefont
  {Sokollik}, \citenamefont {H\"orlein}, \citenamefont {Kiefer}, \citenamefont
  {Jung}, \citenamefont {Schreiber}, \citenamefont {Hegelich}, \citenamefont
  {Yan}, \citenamefont {Meyer-ter Vehn}, \citenamefont {Tajima}, \citenamefont
  {Nickles}, \citenamefont {Sandner},\ and\ \citenamefont {Habs}}]{Henig}%
  \BibitemOpen
  \bibfield  {author} {\bibinfo {author} {\bibfnamefont {A.}~\bibnamefont
  {Henig}}, \bibinfo {author} {\bibfnamefont {S.}~\bibnamefont {Steinke}},
  \bibinfo {author} {\bibfnamefont {M.}~\bibnamefont {Schn\"urer}}, \bibinfo
  {author} {\bibfnamefont {T.}~\bibnamefont {Sokollik}}, \bibinfo {author}
  {\bibfnamefont {R.}~\bibnamefont {H\"orlein}}, \bibinfo {author}
  {\bibfnamefont {D.}~\bibnamefont {Kiefer}}, \bibinfo {author} {\bibfnamefont
  {D.}~\bibnamefont {Jung}}, \bibinfo {author} {\bibfnamefont {J.}~\bibnamefont
  {Schreiber}}, \bibinfo {author} {\bibfnamefont {B.~M.}\ \bibnamefont
  {Hegelich}}, \bibinfo {author} {\bibfnamefont {X.~Q.}\ \bibnamefont {Yan}},
  \bibinfo {author} {\bibfnamefont {J.}~\bibnamefont {Meyer-ter Vehn}},
  \bibinfo {author} {\bibfnamefont {T.}~\bibnamefont {Tajima}}, \bibinfo
  {author} {\bibfnamefont {P.~V.}\ \bibnamefont {Nickles}}, \bibinfo {author}
  {\bibfnamefont {W.}~\bibnamefont {Sandner}}, \ and\ \bibinfo {author}
  {\bibfnamefont {D.}~\bibnamefont {Habs}},\ }\href {\doibase
  10.1103/PhysRevLett.103.245003} {\bibfield  {journal} {\bibinfo  {journal}
  {Phys. Rev. Lett.}\ }\textbf {\bibinfo {volume} {103}},\ \bibinfo {pages}
  {245003} (\bibinfo {year} {2009})}\BibitemShut {NoStop}%
\bibitem [{\citenamefont {Kar}\ \emph {et~al.}(2012)\citenamefont {Kar},
  \citenamefont {Kakolee}, \citenamefont {Qiao}, \citenamefont {Macchi},
  \citenamefont {Cerchez}, \citenamefont {Doria}, \citenamefont {Geissler},
  \citenamefont {McKenna}, \citenamefont {Neely}, \citenamefont {Osterholz},
  \citenamefont {Prasad}, \citenamefont {Quinn}, \citenamefont {Ramakrishna},
  \citenamefont {Sarri}, \citenamefont {Willi}, \citenamefont {Yuan},
  \citenamefont {Zepf},\ and\ \citenamefont {Borghesi}}]{Kar}%
  \BibitemOpen
  \bibfield  {author} {\bibinfo {author} {\bibfnamefont {S.}~\bibnamefont
  {Kar}}, \bibinfo {author} {\bibfnamefont {K.~F.}\ \bibnamefont {Kakolee}},
  \bibinfo {author} {\bibfnamefont {B.}~\bibnamefont {Qiao}}, \bibinfo {author}
  {\bibfnamefont {A.}~\bibnamefont {Macchi}}, \bibinfo {author} {\bibfnamefont
  {M.}~\bibnamefont {Cerchez}}, \bibinfo {author} {\bibfnamefont
  {D.}~\bibnamefont {Doria}}, \bibinfo {author} {\bibfnamefont
  {M.}~\bibnamefont {Geissler}}, \bibinfo {author} {\bibfnamefont
  {P.}~\bibnamefont {McKenna}}, \bibinfo {author} {\bibfnamefont
  {D.}~\bibnamefont {Neely}}, \bibinfo {author} {\bibfnamefont
  {J.}~\bibnamefont {Osterholz}}, \bibinfo {author} {\bibfnamefont
  {R.}~\bibnamefont {Prasad}}, \bibinfo {author} {\bibfnamefont
  {K.}~\bibnamefont {Quinn}}, \bibinfo {author} {\bibfnamefont
  {B.}~\bibnamefont {Ramakrishna}}, \bibinfo {author} {\bibfnamefont
  {G.}~\bibnamefont {Sarri}}, \bibinfo {author} {\bibfnamefont
  {O.}~\bibnamefont {Willi}}, \bibinfo {author} {\bibfnamefont {X.~Y.}\
  \bibnamefont {Yuan}}, \bibinfo {author} {\bibfnamefont {M.}~\bibnamefont
  {Zepf}}, \ and\ \bibinfo {author} {\bibfnamefont {M.}~\bibnamefont
  {Borghesi}},\ }\href {\doibase 10.1103/PhysRevLett.109.185006} {\bibfield
  {journal} {\bibinfo  {journal} {Phys. Rev. Lett.}\ }\textbf {\bibinfo
  {volume} {109}},\ \bibinfo {pages} {185006} (\bibinfo {year}
  {2012})}\BibitemShut {NoStop}%
\bibitem [{\citenamefont {Flippo}\ \emph {et~al.}(2008)\citenamefont {Flippo},
  \citenamefont {d'Humieres}, \citenamefont {Gaillard}, \citenamefont
  {Rassuchine}, \citenamefont {Gautier}, \citenamefont {Schollmeier},
  \citenamefont {Nuernberg}, \citenamefont {Kline}, \citenamefont {Adams},
  \citenamefont {Albright}, \citenamefont {Bakeman}, \citenamefont {Harres},
  \citenamefont {Johnson}, \citenamefont {Korgan}, \citenamefont {Letzring},
  \citenamefont {Malekos}, \citenamefont {Renard-LeGalloudec}, \citenamefont
  {Sentoku}, \citenamefont {Shimada}, \citenamefont {Roth}, \citenamefont
  {Cowan}, \citenamefont {Fernandez},\ and\ \citenamefont {Hegelich}}]{Flippo}%
  \BibitemOpen
  \bibfield  {author} {\bibinfo {author} {\bibfnamefont {K.~A.}\ \bibnamefont
  {Flippo}}, \bibinfo {author} {\bibfnamefont {E.}~\bibnamefont {d'Humieres}},
  \bibinfo {author} {\bibfnamefont {S.~A.}\ \bibnamefont {Gaillard}}, \bibinfo
  {author} {\bibfnamefont {J.}~\bibnamefont {Rassuchine}}, \bibinfo {author}
  {\bibfnamefont {D.~C.}\ \bibnamefont {Gautier}}, \bibinfo {author}
  {\bibfnamefont {M.}~\bibnamefont {Schollmeier}}, \bibinfo {author}
  {\bibfnamefont {F.}~\bibnamefont {Nuernberg}}, \bibinfo {author}
  {\bibfnamefont {J.~L.}\ \bibnamefont {Kline}}, \bibinfo {author}
  {\bibfnamefont {J.}~\bibnamefont {Adams}}, \bibinfo {author} {\bibfnamefont
  {B.}~\bibnamefont {Albright}}, \bibinfo {author} {\bibfnamefont
  {M.}~\bibnamefont {Bakeman}}, \bibinfo {author} {\bibfnamefont
  {K.}~\bibnamefont {Harres}}, \bibinfo {author} {\bibfnamefont {R.~P.}\
  \bibnamefont {Johnson}}, \bibinfo {author} {\bibfnamefont {G.}~\bibnamefont
  {Korgan}}, \bibinfo {author} {\bibfnamefont {S.}~\bibnamefont {Letzring}},
  \bibinfo {author} {\bibfnamefont {S.}~\bibnamefont {Malekos}}, \bibinfo
  {author} {\bibfnamefont {N.}~\bibnamefont {Renard-LeGalloudec}}, \bibinfo
  {author} {\bibfnamefont {Y.}~\bibnamefont {Sentoku}}, \bibinfo {author}
  {\bibfnamefont {T.}~\bibnamefont {Shimada}}, \bibinfo {author} {\bibfnamefont
  {M.}~\bibnamefont {Roth}}, \bibinfo {author} {\bibfnamefont {T.~E.}\
  \bibnamefont {Cowan}}, \bibinfo {author} {\bibfnamefont {J.~C.}\ \bibnamefont
  {Fernandez}}, \ and\ \bibinfo {author} {\bibfnamefont {B.~M.}\ \bibnamefont
  {Hegelich}},\ }\href {\doibase 10.1063/1.2918125} {\bibfield  {journal}
  {\bibinfo  {journal} {Phys. Plasmas}\ }\textbf {\bibinfo {volume} {15}},\
  \bibinfo {pages} {056709} (\bibinfo {year} {2008})}\BibitemShut {NoStop}%
\bibitem [{\citenamefont {Buffechoux}\ \emph {et~al.}(2010)\citenamefont
  {Buffechoux}, \citenamefont {Psikal}, \citenamefont {Nakatsutsumi},
  \citenamefont {Romagnani}, \citenamefont {Andreev}, \citenamefont {Zeil},
  \citenamefont {Amin}, \citenamefont {Antici}, \citenamefont {Burris-Mog},
  \citenamefont {Compant-La-Fontaine}, \citenamefont {d'Humi\`eres},
  \citenamefont {Fourmaux}, \citenamefont {Gaillard}, \citenamefont {Gobet},
  \citenamefont {Hannachi}, \citenamefont {Kraft}, \citenamefont {Mancic},
  \citenamefont {Plaisir}, \citenamefont {Sarri}, \citenamefont {Tarisien},
  \citenamefont {Toncian}, \citenamefont {Schramm}, \citenamefont {Tampo},
  \citenamefont {Audebert}, \citenamefont {Willi}, \citenamefont {Cowan},
  \citenamefont {P\'epin}, \citenamefont {Tikhonchuk}, \citenamefont
  {Borghesi},\ and\ \citenamefont {Fuchs}}]{Buffechoux}%
  \BibitemOpen
  \bibfield  {author} {\bibinfo {author} {\bibfnamefont {S.}~\bibnamefont
  {Buffechoux}}, \bibinfo {author} {\bibfnamefont {J.}~\bibnamefont {Psikal}},
  \bibinfo {author} {\bibfnamefont {M.}~\bibnamefont {Nakatsutsumi}}, \bibinfo
  {author} {\bibfnamefont {L.}~\bibnamefont {Romagnani}}, \bibinfo {author}
  {\bibfnamefont {A.}~\bibnamefont {Andreev}}, \bibinfo {author} {\bibfnamefont
  {K.}~\bibnamefont {Zeil}}, \bibinfo {author} {\bibfnamefont {M.}~\bibnamefont
  {Amin}}, \bibinfo {author} {\bibfnamefont {P.}~\bibnamefont {Antici}},
  \bibinfo {author} {\bibfnamefont {T.}~\bibnamefont {Burris-Mog}}, \bibinfo
  {author} {\bibfnamefont {A.}~\bibnamefont {Compant-La-Fontaine}}, \bibinfo
  {author} {\bibfnamefont {E.}~\bibnamefont {d'Humi\`eres}}, \bibinfo {author}
  {\bibfnamefont {S.}~\bibnamefont {Fourmaux}}, \bibinfo {author}
  {\bibfnamefont {S.}~\bibnamefont {Gaillard}}, \bibinfo {author}
  {\bibfnamefont {F.}~\bibnamefont {Gobet}}, \bibinfo {author} {\bibfnamefont
  {F.}~\bibnamefont {Hannachi}}, \bibinfo {author} {\bibfnamefont
  {S.}~\bibnamefont {Kraft}}, \bibinfo {author} {\bibfnamefont
  {A.}~\bibnamefont {Mancic}}, \bibinfo {author} {\bibfnamefont
  {C.}~\bibnamefont {Plaisir}}, \bibinfo {author} {\bibfnamefont
  {G.}~\bibnamefont {Sarri}}, \bibinfo {author} {\bibfnamefont
  {M.}~\bibnamefont {Tarisien}}, \bibinfo {author} {\bibfnamefont
  {T.}~\bibnamefont {Toncian}}, \bibinfo {author} {\bibfnamefont
  {U.}~\bibnamefont {Schramm}}, \bibinfo {author} {\bibfnamefont
  {M.}~\bibnamefont {Tampo}}, \bibinfo {author} {\bibfnamefont
  {P.}~\bibnamefont {Audebert}}, \bibinfo {author} {\bibfnamefont
  {O.}~\bibnamefont {Willi}}, \bibinfo {author} {\bibfnamefont {T.~E.}\
  \bibnamefont {Cowan}}, \bibinfo {author} {\bibfnamefont {H.}~\bibnamefont
  {P\'epin}}, \bibinfo {author} {\bibfnamefont {V.}~\bibnamefont {Tikhonchuk}},
  \bibinfo {author} {\bibfnamefont {M.}~\bibnamefont {Borghesi}}, \ and\
  \bibinfo {author} {\bibfnamefont {J.}~\bibnamefont {Fuchs}},\ }\href
  {\doibase 10.1103/PhysRevLett.105.015005} {\bibfield  {journal} {\bibinfo
  {journal} {Phys. Rev. Lett.}\ }\textbf {\bibinfo {volume} {105}},\ \bibinfo
  {pages} {015005} (\bibinfo {year} {2010})}\BibitemShut {NoStop}%
\bibitem [{\citenamefont {Burza}\ \emph {et~al.}(2011)\citenamefont {Burza},
  \citenamefont {Gonoskov}, \citenamefont {Genoud}, \citenamefont {Persson},
  \citenamefont {Svensson}, \citenamefont {Quinn}, \citenamefont {McKenna},
  \citenamefont {Marklund},\ and\ \citenamefont {Wahlstrom}}]{Burza}%
  \BibitemOpen
  \bibfield  {author} {\bibinfo {author} {\bibfnamefont {M.}~\bibnamefont
  {Burza}}, \bibinfo {author} {\bibfnamefont {A.}~\bibnamefont {Gonoskov}},
  \bibinfo {author} {\bibfnamefont {G.}~\bibnamefont {Genoud}}, \bibinfo
  {author} {\bibfnamefont {A.}~\bibnamefont {Persson}}, \bibinfo {author}
  {\bibfnamefont {K.}~\bibnamefont {Svensson}}, \bibinfo {author}
  {\bibfnamefont {M.}~\bibnamefont {Quinn}}, \bibinfo {author} {\bibfnamefont
  {P.}~\bibnamefont {McKenna}}, \bibinfo {author} {\bibfnamefont
  {M.}~\bibnamefont {Marklund}}, \ and\ \bibinfo {author} {\bibfnamefont
  {C.-G.}\ \bibnamefont {Wahlstrom}},\ }\href {\doibase
  10.1088/1367-2630/13/1/013030} {\bibfield  {journal} {\bibinfo  {journal}
  {New J. Phys.}\ }\textbf {\bibinfo {volume} {13}},\ \bibinfo {pages} {013030}
  (\bibinfo {year} {2011})}\BibitemShut {NoStop}%
\bibitem [{\citenamefont {Gaillard}\ \emph {et~al.}(2011)\citenamefont
  {Gaillard}, \citenamefont {Kluge}, \citenamefont {Flippo}, \citenamefont
  {Bussmann}, \citenamefont {Gall}, \citenamefont {Lockard}, \citenamefont
  {Geissel}, \citenamefont {Offermann}, \citenamefont {Schollmeier},
  \citenamefont {Sentoku},\ and\ \citenamefont {Cowan}}]{Gaillard}%
  \BibitemOpen
  \bibfield  {author} {\bibinfo {author} {\bibfnamefont {S.~A.}\ \bibnamefont
  {Gaillard}}, \bibinfo {author} {\bibfnamefont {T.}~\bibnamefont {Kluge}},
  \bibinfo {author} {\bibfnamefont {K.~A.}\ \bibnamefont {Flippo}}, \bibinfo
  {author} {\bibfnamefont {M.}~\bibnamefont {Bussmann}}, \bibinfo {author}
  {\bibfnamefont {B.}~\bibnamefont {Gall}}, \bibinfo {author} {\bibfnamefont
  {T.}~\bibnamefont {Lockard}}, \bibinfo {author} {\bibfnamefont
  {M.}~\bibnamefont {Geissel}}, \bibinfo {author} {\bibfnamefont {D.~T.}\
  \bibnamefont {Offermann}}, \bibinfo {author} {\bibfnamefont {M.}~\bibnamefont
  {Schollmeier}}, \bibinfo {author} {\bibfnamefont {Y.}~\bibnamefont
  {Sentoku}}, \ and\ \bibinfo {author} {\bibfnamefont {T.~E.}\ \bibnamefont
  {Cowan}},\ }\href {\doibase 10.1063/1.3575624} {\bibfield  {journal}
  {\bibinfo  {journal} {Phys. Plasmas}\ }\textbf {\bibinfo {volume} {18}},\
  \bibinfo {pages} {056710} (\bibinfo {year} {2011})}\BibitemShut {NoStop}%
\bibitem [{\citenamefont {Markey}\ \emph {et~al.}(2010)\citenamefont {Markey},
  \citenamefont {McKenna}, \citenamefont {Brenner}, \citenamefont {Carroll},
  \citenamefont {G\"unther}, \citenamefont {Harres}, \citenamefont {Kar},
  \citenamefont {Lancaster}, \citenamefont {N\"urnberg}, \citenamefont {Quinn},
  \citenamefont {Robinson}, \citenamefont {Roth}, \citenamefont {Zepf},\ and\
  \citenamefont {Neely}}]{Markey}%
  \BibitemOpen
  \bibfield  {author} {\bibinfo {author} {\bibfnamefont {K.}~\bibnamefont
  {Markey}}, \bibinfo {author} {\bibfnamefont {P.}~\bibnamefont {McKenna}},
  \bibinfo {author} {\bibfnamefont {C.~M.}\ \bibnamefont {Brenner}}, \bibinfo
  {author} {\bibfnamefont {D.~C.}\ \bibnamefont {Carroll}}, \bibinfo {author}
  {\bibfnamefont {M.~M.}\ \bibnamefont {G\"unther}}, \bibinfo {author}
  {\bibfnamefont {K.}~\bibnamefont {Harres}}, \bibinfo {author} {\bibfnamefont
  {S.}~\bibnamefont {Kar}}, \bibinfo {author} {\bibfnamefont {K.}~\bibnamefont
  {Lancaster}}, \bibinfo {author} {\bibfnamefont {F.}~\bibnamefont
  {N\"urnberg}}, \bibinfo {author} {\bibfnamefont {M.~N.}\ \bibnamefont
  {Quinn}}, \bibinfo {author} {\bibfnamefont {A.~P.~L.}\ \bibnamefont
  {Robinson}}, \bibinfo {author} {\bibfnamefont {M.}~\bibnamefont {Roth}},
  \bibinfo {author} {\bibfnamefont {M.}~\bibnamefont {Zepf}}, \ and\ \bibinfo
  {author} {\bibfnamefont {D.}~\bibnamefont {Neely}},\ }\href {\doibase
  10.1103/PhysRevLett.105.195008} {\bibfield  {journal} {\bibinfo  {journal}
  {Phys. Rev. Lett.}\ }\textbf {\bibinfo {volume} {105}},\ \bibinfo {pages}
  {195008} (\bibinfo {year} {2010})}\BibitemShut {NoStop}%
\bibitem [{\citenamefont {Pfotenhauer}\ \emph {et~al.}(2010)\citenamefont
  {Pfotenhauer}, \citenamefont {Jaeckel}, \citenamefont {Polz}, \citenamefont
  {Steinke}, \citenamefont {Schlenvoigt}, \citenamefont {Heymann},
  \citenamefont {Robinson},\ and\ \citenamefont {Kaluza}}]{Pfotenhauer}%
  \BibitemOpen
  \bibfield  {author} {\bibinfo {author} {\bibfnamefont {S.~M.}\ \bibnamefont
  {Pfotenhauer}}, \bibinfo {author} {\bibfnamefont {O.}~\bibnamefont
  {Jaeckel}}, \bibinfo {author} {\bibfnamefont {J.}~\bibnamefont {Polz}},
  \bibinfo {author} {\bibfnamefont {S.}~\bibnamefont {Steinke}}, \bibinfo
  {author} {\bibfnamefont {H.-P.}\ \bibnamefont {Schlenvoigt}}, \bibinfo
  {author} {\bibfnamefont {J.}~\bibnamefont {Heymann}}, \bibinfo {author}
  {\bibfnamefont {A.~P.~L.}\ \bibnamefont {Robinson}}, \ and\ \bibinfo {author}
  {\bibfnamefont {M.~C.}\ \bibnamefont {Kaluza}},\ }\href {\doibase
  10.1088/1367-2630/12/10/103009} {\bibfield  {journal} {\bibinfo  {journal}
  {New J. Phys.}\ }\textbf {\bibinfo {volume} {12}},\ \bibinfo {pages} {103009}
  (\bibinfo {year} {2010})}\BibitemShut {NoStop}%
\bibitem [{\citenamefont {Dalui}\ \emph {et~al.}(2017)\citenamefont {Dalui},
  \citenamefont {Kundu}, \citenamefont {Sarkar}, \citenamefont {Tata},
  \citenamefont {Pasley}, \citenamefont {Ayyub},\ and\ \citenamefont
  {Krishnamurthy}}]{Dalui}%
  \BibitemOpen
  \bibfield  {author} {\bibinfo {author} {\bibfnamefont {M.}~\bibnamefont
  {Dalui}}, \bibinfo {author} {\bibfnamefont {M.}~\bibnamefont {Kundu}},
  \bibinfo {author} {\bibfnamefont {S.}~\bibnamefont {Sarkar}}, \bibinfo
  {author} {\bibfnamefont {S.}~\bibnamefont {Tata}}, \bibinfo {author}
  {\bibfnamefont {J.}~\bibnamefont {Pasley}}, \bibinfo {author} {\bibfnamefont
  {P.}~\bibnamefont {Ayyub}}, \ and\ \bibinfo {author} {\bibfnamefont
  {M.}~\bibnamefont {Krishnamurthy}},\ }\href {\doibase 10.1063/1.4973887}
  {\bibfield  {journal} {\bibinfo  {journal} {Physics of Plasmas}\ }\textbf
  {\bibinfo {volume} {24}},\ \bibinfo {pages} {010703} (\bibinfo {year}
  {2017})}\BibitemShut {NoStop}%
\bibitem [{\citenamefont {Floquet}\ \emph {et~al.}(2013)\citenamefont
  {Floquet}, \citenamefont {Klimo}, \citenamefont {Psikal}, \citenamefont
  {Velyhan}, \citenamefont {Limpouch}, \citenamefont {Proska}, \citenamefont
  {Novotny}, \citenamefont {Stolcova}, \citenamefont {Macchi}, \citenamefont
  {Sgattoni}, \citenamefont {Vassura}, \citenamefont {Labate}, \citenamefont
  {Baffigi}, \citenamefont {Gizzi}, \citenamefont {Martin},\ and\ \citenamefont
  {Ceccotti}}]{Floquet}%
  \BibitemOpen
  \bibfield  {author} {\bibinfo {author} {\bibfnamefont {V.}~\bibnamefont
  {Floquet}}, \bibinfo {author} {\bibfnamefont {O.}~\bibnamefont {Klimo}},
  \bibinfo {author} {\bibfnamefont {J.}~\bibnamefont {Psikal}}, \bibinfo
  {author} {\bibfnamefont {A.}~\bibnamefont {Velyhan}}, \bibinfo {author}
  {\bibfnamefont {J.}~\bibnamefont {Limpouch}}, \bibinfo {author}
  {\bibfnamefont {J.}~\bibnamefont {Proska}}, \bibinfo {author} {\bibfnamefont
  {F.}~\bibnamefont {Novotny}}, \bibinfo {author} {\bibfnamefont
  {L.}~\bibnamefont {Stolcova}}, \bibinfo {author} {\bibfnamefont
  {A.}~\bibnamefont {Macchi}}, \bibinfo {author} {\bibfnamefont
  {A.}~\bibnamefont {Sgattoni}}, \bibinfo {author} {\bibfnamefont
  {L.}~\bibnamefont {Vassura}}, \bibinfo {author} {\bibfnamefont
  {L.}~\bibnamefont {Labate}}, \bibinfo {author} {\bibfnamefont
  {F.}~\bibnamefont {Baffigi}}, \bibinfo {author} {\bibfnamefont {L.~A.}\
  \bibnamefont {Gizzi}}, \bibinfo {author} {\bibfnamefont {P.}~\bibnamefont
  {Martin}}, \ and\ \bibinfo {author} {\bibfnamefont {T.}~\bibnamefont
  {Ceccotti}},\ }\href {\doibase 10.1063/1.4819239} {\bibfield  {journal}
  {\bibinfo  {journal} {Journal of Applied Physics}\ }\textbf {\bibinfo
  {volume} {114}},\ \bibinfo {pages} {083305} (\bibinfo {year}
  {2013})}\BibitemShut {NoStop}%
\bibitem [{\citenamefont {Jiang}\ \emph {et~al.}(2016)\citenamefont {Jiang},
  \citenamefont {Ji}, \citenamefont {Audesirk}, \citenamefont {George},
  \citenamefont {Snyder}, \citenamefont {Krygier}, \citenamefont {Poole},
  \citenamefont {Willis}, \citenamefont {Daskalova}, \citenamefont {Chowdhury},
  \citenamefont {Lewis}, \citenamefont {Schumacher}, \citenamefont {Pukhov},
  \citenamefont {Freeman},\ and\ \citenamefont {Akli}}]{Jiang}%
  \BibitemOpen
  \bibfield  {author} {\bibinfo {author} {\bibfnamefont {S.}~\bibnamefont
  {Jiang}}, \bibinfo {author} {\bibfnamefont {L.~L.}\ \bibnamefont {Ji}},
  \bibinfo {author} {\bibfnamefont {H.}~\bibnamefont {Audesirk}}, \bibinfo
  {author} {\bibfnamefont {K.~M.}\ \bibnamefont {George}}, \bibinfo {author}
  {\bibfnamefont {J.}~\bibnamefont {Snyder}}, \bibinfo {author} {\bibfnamefont
  {A.}~\bibnamefont {Krygier}}, \bibinfo {author} {\bibfnamefont
  {P.}~\bibnamefont {Poole}}, \bibinfo {author} {\bibfnamefont
  {C.}~\bibnamefont {Willis}}, \bibinfo {author} {\bibfnamefont
  {R.}~\bibnamefont {Daskalova}}, \bibinfo {author} {\bibfnamefont
  {E.}~\bibnamefont {Chowdhury}}, \bibinfo {author} {\bibfnamefont {N.~S.}\
  \bibnamefont {Lewis}}, \bibinfo {author} {\bibfnamefont {D.~W.}\ \bibnamefont
  {Schumacher}}, \bibinfo {author} {\bibfnamefont {A.}~\bibnamefont {Pukhov}},
  \bibinfo {author} {\bibfnamefont {R.~R.}\ \bibnamefont {Freeman}}, \ and\
  \bibinfo {author} {\bibfnamefont {K.~U.}\ \bibnamefont {Akli}},\ }\href
  {\doibase 10.1103/PhysRevLett.116.085002} {\bibfield  {journal} {\bibinfo
  {journal} {Phys. Rev. Lett.}\ }\textbf {\bibinfo {volume} {116}},\ \bibinfo
  {pages} {085002} (\bibinfo {year} {2016})}\BibitemShut {NoStop}%
\bibitem [{\citenamefont {Margarone}\ \emph {et~al.}(2015)\citenamefont
  {Margarone}, \citenamefont {Kim}, \citenamefont {Psikal}, \citenamefont
  {Kaufman}, \citenamefont {Mocek}, \citenamefont {Choi}, \citenamefont
  {Stolcova}, \citenamefont {Proska}, \citenamefont {Choukourov}, \citenamefont
  {Melnichuk}, \citenamefont {Klimo}, \citenamefont {Limpouch}, \citenamefont
  {Sung}, \citenamefont {Lee}, \citenamefont {Korn},\ and\ \citenamefont
  {Jeong}}]{Margarone1}%
  \BibitemOpen
  \bibfield  {author} {\bibinfo {author} {\bibfnamefont {D.}~\bibnamefont
  {Margarone}}, \bibinfo {author} {\bibfnamefont {I.~J.}\ \bibnamefont {Kim}},
  \bibinfo {author} {\bibfnamefont {J.}~\bibnamefont {Psikal}}, \bibinfo
  {author} {\bibfnamefont {J.}~\bibnamefont {Kaufman}}, \bibinfo {author}
  {\bibfnamefont {T.}~\bibnamefont {Mocek}}, \bibinfo {author} {\bibfnamefont
  {I.~W.}\ \bibnamefont {Choi}}, \bibinfo {author} {\bibfnamefont
  {L.}~\bibnamefont {Stolcova}}, \bibinfo {author} {\bibfnamefont
  {J.}~\bibnamefont {Proska}}, \bibinfo {author} {\bibfnamefont
  {A.}~\bibnamefont {Choukourov}}, \bibinfo {author} {\bibfnamefont
  {I.}~\bibnamefont {Melnichuk}}, \bibinfo {author} {\bibfnamefont
  {O.}~\bibnamefont {Klimo}}, \bibinfo {author} {\bibfnamefont
  {J.}~\bibnamefont {Limpouch}}, \bibinfo {author} {\bibfnamefont {J.~H.}\
  \bibnamefont {Sung}}, \bibinfo {author} {\bibfnamefont {S.~K.}\ \bibnamefont
  {Lee}}, \bibinfo {author} {\bibfnamefont {G.}~\bibnamefont {Korn}}, \ and\
  \bibinfo {author} {\bibfnamefont {T.~M.}\ \bibnamefont {Jeong}},\ }\href
  {\doibase 10.1103/PhysRevSTAB.18.071304} {\bibfield  {journal} {\bibinfo
  {journal} {Phys. Rev. ST Accel. Beams}\ }\textbf {\bibinfo {volume} {18}},\
  \bibinfo {pages} {071304} (\bibinfo {year} {2015})}\BibitemShut {NoStop}%
\bibitem [{\citenamefont {Passoni}\ \emph {et~al.}(2016)\citenamefont
  {Passoni}, \citenamefont {Sgattoni}, \citenamefont {Prencipe}, \citenamefont
  {Fedeli}, \citenamefont {Dellasega}, \citenamefont {Cialfi}, \citenamefont
  {Choi}, \citenamefont {Kim}, \citenamefont {Janulewicz}, \citenamefont {Lee},
  \citenamefont {Sung}, \citenamefont {Lee},\ and\ \citenamefont
  {Nam}}]{Passoni2}%
  \BibitemOpen
  \bibfield  {author} {\bibinfo {author} {\bibfnamefont {M.}~\bibnamefont
  {Passoni}}, \bibinfo {author} {\bibfnamefont {A.}~\bibnamefont {Sgattoni}},
  \bibinfo {author} {\bibfnamefont {I.}~\bibnamefont {Prencipe}}, \bibinfo
  {author} {\bibfnamefont {L.}~\bibnamefont {Fedeli}}, \bibinfo {author}
  {\bibfnamefont {D.}~\bibnamefont {Dellasega}}, \bibinfo {author}
  {\bibfnamefont {L.}~\bibnamefont {Cialfi}}, \bibinfo {author} {\bibfnamefont
  {I.~W.}\ \bibnamefont {Choi}}, \bibinfo {author} {\bibfnamefont {I.~J.}\
  \bibnamefont {Kim}}, \bibinfo {author} {\bibfnamefont {K.~A.}\ \bibnamefont
  {Janulewicz}}, \bibinfo {author} {\bibfnamefont {H.~W.}\ \bibnamefont {Lee}},
  \bibinfo {author} {\bibfnamefont {J.~H.}\ \bibnamefont {Sung}}, \bibinfo
  {author} {\bibfnamefont {S.~K.}\ \bibnamefont {Lee}}, \ and\ \bibinfo
  {author} {\bibfnamefont {C.~H.}\ \bibnamefont {Nam}},\ }\href {\doibase
  10.1103/PhysRevAccelBeams.19.061301} {\bibfield  {journal} {\bibinfo
  {journal} {Phys. Rev. Accel. Beams}\ }\textbf {\bibinfo {volume} {19}},\
  \bibinfo {pages} {061301} (\bibinfo {year} {2016})}\BibitemShut {NoStop}%
\bibitem [{\citenamefont {Zou}\ \emph {et~al.}(2017)\citenamefont {Zou},
  \citenamefont {Pukhov}, \citenamefont {Yi}, \citenamefont {Zhuo},
  \citenamefont {Yu}, \citenamefont {Yin},\ and\ \citenamefont {Shao}}]{Zou}%
  \BibitemOpen
  \bibfield  {author} {\bibinfo {author} {\bibfnamefont {D.~B.}\ \bibnamefont
  {Zou}}, \bibinfo {author} {\bibfnamefont {A.}~\bibnamefont {Pukhov}},
  \bibinfo {author} {\bibfnamefont {L.~Q.}\ \bibnamefont {Yi}}, \bibinfo
  {author} {\bibfnamefont {H.~B.}\ \bibnamefont {Zhuo}}, \bibinfo {author}
  {\bibfnamefont {T.~P.}\ \bibnamefont {Yu}}, \bibinfo {author} {\bibfnamefont
  {Y.}~\bibnamefont {Yin}}, \ and\ \bibinfo {author} {\bibfnamefont {F.~Q.}\
  \bibnamefont {Shao}},\ }\href {\doibase 10.1038/srep42666} {\bibfield
  {journal} {\bibinfo  {journal} {Sci. Rep.}\ }\textbf {\bibinfo {volume}
  {7}},\ \bibinfo {pages} {42666} (\bibinfo {year} {2017})}\BibitemShut
  {NoStop}%
\bibitem [{\citenamefont {L\"ubcke}\ \emph {et~al.}(2017)\citenamefont
  {L\"ubcke}, \citenamefont {Andreev}, \citenamefont {H\"ohm}, \citenamefont
  {Grunwald}, \citenamefont {Ehrentraut},\ and\ \citenamefont
  {Schn\"urer}}]{Lubcke}%
  \BibitemOpen
  \bibfield  {author} {\bibinfo {author} {\bibfnamefont {A.}~\bibnamefont
  {L\"ubcke}}, \bibinfo {author} {\bibfnamefont {A.~A.}\ \bibnamefont
  {Andreev}}, \bibinfo {author} {\bibfnamefont {S.}~\bibnamefont {H\"ohm}},
  \bibinfo {author} {\bibfnamefont {R.}~\bibnamefont {Grunwald}}, \bibinfo
  {author} {\bibfnamefont {L.}~\bibnamefont {Ehrentraut}}, \ and\ \bibinfo
  {author} {\bibfnamefont {M.}~\bibnamefont {Schn\"urer}},\ }\href {\doibase
  10.1038/srep44030} {\bibfield  {journal} {\bibinfo  {journal} {Sci. Rep.}\
  }\textbf {\bibinfo {volume} {7}},\ \bibinfo {pages} {44030} (\bibinfo {year}
  {2017})}\BibitemShut {NoStop}%
\bibitem [{\citenamefont {Nodera}\ \emph {et~al.}(2008)\citenamefont {Nodera},
  \citenamefont {Kawata}, \citenamefont {Onuma}, \citenamefont {Limpouch},
  \citenamefont {Klimo},\ and\ \citenamefont {Kikuchi}}]{Nodera}%
  \BibitemOpen
  \bibfield  {author} {\bibinfo {author} {\bibfnamefont {Y.}~\bibnamefont
  {Nodera}}, \bibinfo {author} {\bibfnamefont {S.}~\bibnamefont {Kawata}},
  \bibinfo {author} {\bibfnamefont {N.}~\bibnamefont {Onuma}}, \bibinfo
  {author} {\bibfnamefont {J.}~\bibnamefont {Limpouch}}, \bibinfo {author}
  {\bibfnamefont {O.}~\bibnamefont {Klimo}}, \ and\ \bibinfo {author}
  {\bibfnamefont {T.}~\bibnamefont {Kikuchi}},\ }\href {\doibase
  10.1103/PhysRevE.78.046401} {\bibfield  {journal} {\bibinfo  {journal} {Phys.
  Rev. E}\ }\textbf {\bibinfo {volume} {78}},\ \bibinfo {pages} {046401}
  (\bibinfo {year} {2008})}\BibitemShut {NoStop}%
\bibitem [{\citenamefont {Cao}\ \emph {et~al.}(2010)\citenamefont {Cao},
  \citenamefont {Gu}, \citenamefont {Zhao}, \citenamefont {Cao}, \citenamefont
  {Huang}, \citenamefont {Zhou}, \citenamefont {He}, \citenamefont {Yu},\ and\
  \citenamefont {Yu}}]{Cao}%
  \BibitemOpen
  \bibfield  {author} {\bibinfo {author} {\bibfnamefont {L.}~\bibnamefont
  {Cao}}, \bibinfo {author} {\bibfnamefont {Y.}~\bibnamefont {Gu}}, \bibinfo
  {author} {\bibfnamefont {Z.}~\bibnamefont {Zhao}}, \bibinfo {author}
  {\bibfnamefont {L.}~\bibnamefont {Cao}}, \bibinfo {author} {\bibfnamefont
  {W.}~\bibnamefont {Huang}}, \bibinfo {author} {\bibfnamefont
  {W.}~\bibnamefont {Zhou}}, \bibinfo {author} {\bibfnamefont {X.~T.}\
  \bibnamefont {He}}, \bibinfo {author} {\bibfnamefont {W.}~\bibnamefont {Yu}},
  \ and\ \bibinfo {author} {\bibfnamefont {M.~Y.}\ \bibnamefont {Yu}},\ }\href
  {\doibase 10.1063/1.3360298} {\bibfield  {journal} {\bibinfo  {journal}
  {Physics of Plasmas}\ }\textbf {\bibinfo {volume} {17}},\ \bibinfo {pages}
  {043103} (\bibinfo {year} {2010})}\BibitemShut {NoStop}%
\bibitem [{\citenamefont {Klimo}\ \emph {et~al.}(2011)\citenamefont {Klimo},
  \citenamefont {Psikal}, \citenamefont {Limpouch}, \citenamefont {Proska},
  \citenamefont {Novotny}, \citenamefont {Ceccotti}, \citenamefont {Floquet},\
  and\ \citenamefont {Kawata}}]{Klimo}%
  \BibitemOpen
  \bibfield  {author} {\bibinfo {author} {\bibfnamefont {O.}~\bibnamefont
  {Klimo}}, \bibinfo {author} {\bibfnamefont {J.}~\bibnamefont {Psikal}},
  \bibinfo {author} {\bibfnamefont {J.}~\bibnamefont {Limpouch}}, \bibinfo
  {author} {\bibfnamefont {J.}~\bibnamefont {Proska}}, \bibinfo {author}
  {\bibfnamefont {F.}~\bibnamefont {Novotny}}, \bibinfo {author} {\bibfnamefont
  {T.}~\bibnamefont {Ceccotti}}, \bibinfo {author} {\bibfnamefont
  {V.}~\bibnamefont {Floquet}}, \ and\ \bibinfo {author} {\bibfnamefont
  {S.}~\bibnamefont {Kawata}},\ }\href
  {http://stacks.iop.org/1367-2630/13/i=5/a=053028} {\bibfield  {journal}
  {\bibinfo  {journal} {New Journal of Physics}\ }\textbf {\bibinfo {volume}
  {13}},\ \bibinfo {pages} {053028} (\bibinfo {year} {2011})}\BibitemShut
  {NoStop}%
\bibitem [{\citenamefont {Margarone}\ \emph {et~al.}(2012)\citenamefont
  {Margarone}, \citenamefont {Klimo}, \citenamefont {Kim}, \citenamefont
  {Prok\ifmmode~\mathring{u}\else \r{u}\fi{}pek}, \citenamefont {Limpouch},
  \citenamefont {Jeong}, \citenamefont {Mocek}, \citenamefont
  {P\ifmmode~\check{s}\else \v{s}\fi{}ikal}, \citenamefont {Kim}, \citenamefont
  {Pro\ifmmode~\check{s}\else \v{s}\fi{}ka}, \citenamefont {Nam}, \citenamefont
  {\ifmmode~\check{S}\else \v{S}\fi{}tolcov\'a}, \citenamefont {Choi},
  \citenamefont {Lee}, \citenamefont {Sung}, \citenamefont {Yu},\ and\
  \citenamefont {Korn}}]{Margarone2}%
  \BibitemOpen
  \bibfield  {author} {\bibinfo {author} {\bibfnamefont {D.}~\bibnamefont
  {Margarone}}, \bibinfo {author} {\bibfnamefont {O.}~\bibnamefont {Klimo}},
  \bibinfo {author} {\bibfnamefont {I.~J.}\ \bibnamefont {Kim}}, \bibinfo
  {author} {\bibfnamefont {J.}~\bibnamefont {Prok\ifmmode~\mathring{u}\else
  \r{u}\fi{}pek}}, \bibinfo {author} {\bibfnamefont {J.}~\bibnamefont
  {Limpouch}}, \bibinfo {author} {\bibfnamefont {T.~M.}\ \bibnamefont {Jeong}},
  \bibinfo {author} {\bibfnamefont {T.}~\bibnamefont {Mocek}}, \bibinfo
  {author} {\bibfnamefont {J.}~\bibnamefont {P\ifmmode~\check{s}\else
  \v{s}\fi{}ikal}}, \bibinfo {author} {\bibfnamefont {H.~T.}\ \bibnamefont
  {Kim}}, \bibinfo {author} {\bibfnamefont {J.}~\bibnamefont
  {Pro\ifmmode~\check{s}\else \v{s}\fi{}ka}}, \bibinfo {author} {\bibfnamefont
  {K.~H.}\ \bibnamefont {Nam}}, \bibinfo {author} {\bibfnamefont
  {L.}~\bibnamefont {\ifmmode~\check{S}\else \v{S}\fi{}tolcov\'a}}, \bibinfo
  {author} {\bibfnamefont {I.~W.}\ \bibnamefont {Choi}}, \bibinfo {author}
  {\bibfnamefont {S.~K.}\ \bibnamefont {Lee}}, \bibinfo {author} {\bibfnamefont
  {J.~H.}\ \bibnamefont {Sung}}, \bibinfo {author} {\bibfnamefont {T.~J.}\
  \bibnamefont {Yu}}, \ and\ \bibinfo {author} {\bibfnamefont {G.}~\bibnamefont
  {Korn}},\ }\href {\doibase 10.1103/PhysRevLett.109.234801} {\bibfield
  {journal} {\bibinfo  {journal} {Phys. Rev. Lett.}\ }\textbf {\bibinfo
  {volume} {109}},\ \bibinfo {pages} {234801} (\bibinfo {year}
  {2012})}\BibitemShut {NoStop}%
\bibitem [{\citenamefont {Andreev}\ \emph {et~al.}(2011)\citenamefont
  {Andreev}, \citenamefont {Kumar}, \citenamefont {Platonov},\ and\
  \citenamefont {Pukhov}}]{Andreev1}%
  \BibitemOpen
  \bibfield  {author} {\bibinfo {author} {\bibfnamefont {A.}~\bibnamefont
  {Andreev}}, \bibinfo {author} {\bibfnamefont {N.}~\bibnamefont {Kumar}},
  \bibinfo {author} {\bibfnamefont {K.}~\bibnamefont {Platonov}}, \ and\
  \bibinfo {author} {\bibfnamefont {A.}~\bibnamefont {Pukhov}},\ }\href
  {\doibase 10.1063/1.3641965} {\bibfield  {journal} {\bibinfo  {journal}
  {Physics of Plasmas}\ }\textbf {\bibinfo {volume} {18}},\ \bibinfo {pages}
  {103103} (\bibinfo {year} {2011})}\BibitemShut {NoStop}%
\bibitem [{\citenamefont {Blanco}\ \emph {et~al.}(2017)\citenamefont {Blanco},
  \citenamefont {Flores-Arias}, \citenamefont {Ruiz},\ and\ \citenamefont
  {Vranic}}]{Blanco}%
  \BibitemOpen
  \bibfield  {author} {\bibinfo {author} {\bibfnamefont {M.}~\bibnamefont
  {Blanco}}, \bibinfo {author} {\bibfnamefont {M.~T.}\ \bibnamefont
  {Flores-Arias}}, \bibinfo {author} {\bibfnamefont {C.}~\bibnamefont {Ruiz}},
  \ and\ \bibinfo {author} {\bibfnamefont {M.}~\bibnamefont {Vranic}},\ }\href
  {http://stacks.iop.org/1367-2630/19/i=3/a=033004} {\bibfield  {journal}
  {\bibinfo  {journal} {New Journal of Physics}\ }\textbf {\bibinfo {volume}
  {19}},\ \bibinfo {pages} {033004} (\bibinfo {year} {2017})}\BibitemShut
  {NoStop}%
\bibitem [{\citenamefont {Andreev}\ \emph {et~al.}(2016)\citenamefont
  {Andreev}, \citenamefont {Platonov}, \citenamefont {Braenzel}, \citenamefont
  {LÃŒbcke}, \citenamefont {Das}, \citenamefont {Messaoudi}, \citenamefont
  {Grunwald}, \citenamefont {Gray}, \citenamefont {McGlynn},\ and\
  \citenamefont {SchnÃŒrer}}]{Andreev2}%
  \BibitemOpen
  \bibfield  {author} {\bibinfo {author} {\bibfnamefont {A.}~\bibnamefont
  {Andreev}}, \bibinfo {author} {\bibfnamefont {K.}~\bibnamefont {Platonov}},
  \bibinfo {author} {\bibfnamefont {J.}~\bibnamefont {Braenzel}}, \bibinfo
  {author} {\bibfnamefont {A.}~\bibnamefont {LÃŒbcke}}, \bibinfo {author}
  {\bibfnamefont {S.}~\bibnamefont {Das}}, \bibinfo {author} {\bibfnamefont
  {H.}~\bibnamefont {Messaoudi}}, \bibinfo {author} {\bibfnamefont
  {R.}~\bibnamefont {Grunwald}}, \bibinfo {author} {\bibfnamefont
  {C.}~\bibnamefont {Gray}}, \bibinfo {author} {\bibfnamefont {E.}~\bibnamefont
  {McGlynn}}, \ and\ \bibinfo {author} {\bibfnamefont {M.}~\bibnamefont
  {SchnÃŒrer}},\ }\href {http://stacks.iop.org/0741-3335/58/i=1/a=014038}
  {\bibfield  {journal} {\bibinfo  {journal} {Plasma Physics and Controlled
  Fusion}\ }\textbf {\bibinfo {volume} {58}},\ \bibinfo {pages} {014038}
  (\bibinfo {year} {2016})}\BibitemShut {NoStop}%
\bibitem [{\citenamefont {Magnusson}\ \emph {et~al.}(2017)\citenamefont
  {Magnusson}, \citenamefont {Gonoskov},\ and\ \citenamefont
  {Marklund}}]{Magnusson}%
  \BibitemOpen
  \bibfield  {author} {\bibinfo {author} {\bibfnamefont {J.}~\bibnamefont
  {Magnusson}}, \bibinfo {author} {\bibfnamefont {A.}~\bibnamefont {Gonoskov}},
  \ and\ \bibinfo {author} {\bibfnamefont {M.}~\bibnamefont {Marklund}},\
  }\href {\doibase 10.1140/epjd/e2017-80228-1} {\bibfield  {journal} {\bibinfo
  {journal} {Eur. Phys. J. D}\ }\textbf {\bibinfo {volume} {71}},\ \bibinfo
  {pages} {231} (\bibinfo {year} {2017})}\BibitemShut {NoStop}%
\bibitem [{\citenamefont {Mackenroth}\ \emph {et~al.}(2017)\citenamefont
  {Mackenroth}, \citenamefont {Gonoskov},\ and\ \citenamefont
  {Marklund}}]{mackenroth.epjd.2017}%
  \BibitemOpen
  \bibfield  {author} {\bibinfo {author} {\bibfnamefont {F.}~\bibnamefont
  {Mackenroth}}, \bibinfo {author} {\bibfnamefont {A.}~\bibnamefont
  {Gonoskov}}, \ and\ \bibinfo {author} {\bibfnamefont {M.}~\bibnamefont
  {Marklund}},\ }\href {\doibase 10.1140/epjd/e2017-80184-8} {\bibfield
  {journal} {\bibinfo  {journal} {The European Physical Journal D}\ }\textbf
  {\bibinfo {volume} {71}},\ \bibinfo {pages} {204} (\bibinfo {year}
  {2017})}\BibitemShut {NoStop}%
\bibitem [{\citenamefont {Kar}\ \emph {et~al.}(2016)\citenamefont {Kar},
  \citenamefont {Ahmed}, \citenamefont {Prasad}, \citenamefont {Cerchez},
  \citenamefont {Brauckmann}, \citenamefont {Aurand}, \citenamefont {Cantono},
  \citenamefont {Hadjisolomou}, \citenamefont {Lewis}, \citenamefont {Macchi}
  \emph {et~al.}}]{kar.ncom.2016}%
  \BibitemOpen
  \bibfield  {author} {\bibinfo {author} {\bibfnamefont {S.}~\bibnamefont
  {Kar}}, \bibinfo {author} {\bibfnamefont {H.}~\bibnamefont {Ahmed}}, \bibinfo
  {author} {\bibfnamefont {R.}~\bibnamefont {Prasad}}, \bibinfo {author}
  {\bibfnamefont {M.}~\bibnamefont {Cerchez}}, \bibinfo {author} {\bibfnamefont
  {S.}~\bibnamefont {Brauckmann}}, \bibinfo {author} {\bibfnamefont
  {B.}~\bibnamefont {Aurand}}, \bibinfo {author} {\bibfnamefont
  {G.}~\bibnamefont {Cantono}}, \bibinfo {author} {\bibfnamefont
  {P.}~\bibnamefont {Hadjisolomou}}, \bibinfo {author} {\bibfnamefont {C.~L.}\
  \bibnamefont {Lewis}}, \bibinfo {author} {\bibfnamefont {A.}~\bibnamefont
  {Macchi}},  \emph {et~al.},\ }\href {http://dx.doi.org/10.1038/ncomms10792}
  {\bibfield  {journal} {\bibinfo  {journal} {Nature communications}\ }\textbf
  {\bibinfo {volume} {7}} (\bibinfo {year} {2016})}\BibitemShut {NoStop}%
\bibitem [{\citenamefont {Gonzalez-Izquierdo}\ \emph
  {et~al.}(2016)\citenamefont {Gonzalez-Izquierdo}, \citenamefont {King},
  \citenamefont {Gray}, \citenamefont {Wilson}, \citenamefont {Dance},
  \citenamefont {Powell}, \citenamefont {Maclellan}, \citenamefont {McCreadie},
  \citenamefont {Butler}, \citenamefont {Hawkes} \emph
  {et~al.}}]{gonzalez.ncom.2016}%
  \BibitemOpen
  \bibfield  {author} {\bibinfo {author} {\bibfnamefont {B.}~\bibnamefont
  {Gonzalez-Izquierdo}}, \bibinfo {author} {\bibfnamefont {M.}~\bibnamefont
  {King}}, \bibinfo {author} {\bibfnamefont {R.~J.}\ \bibnamefont {Gray}},
  \bibinfo {author} {\bibfnamefont {R.}~\bibnamefont {Wilson}}, \bibinfo
  {author} {\bibfnamefont {R.~J.}\ \bibnamefont {Dance}}, \bibinfo {author}
  {\bibfnamefont {H.}~\bibnamefont {Powell}}, \bibinfo {author} {\bibfnamefont
  {D.~A.}\ \bibnamefont {Maclellan}}, \bibinfo {author} {\bibfnamefont
  {J.}~\bibnamefont {McCreadie}}, \bibinfo {author} {\bibfnamefont {N.~M.}\
  \bibnamefont {Butler}}, \bibinfo {author} {\bibfnamefont {S.}~\bibnamefont
  {Hawkes}},  \emph {et~al.},\ }\href {\doibase 10.1038/ncomms12891} {\bibfield
   {journal} {\bibinfo  {journal} {Nature communications}\ }\textbf {\bibinfo
  {volume} {7}},\ \bibinfo {pages} {12891} (\bibinfo {year}
  {2016})}\BibitemShut {NoStop}%
\bibitem [{\citenamefont {Giuffrida}\ \emph {et~al.}(2017)\citenamefont
  {Giuffrida}, \citenamefont {Svensson}, \citenamefont {Psikal}, \citenamefont
  {Dalui}, \citenamefont {Ekerfelt}, \citenamefont {Gallardo~Gonzalez},
  \citenamefont {Lundh}, \citenamefont {Persson}, \citenamefont {Lutoslawski},
  \citenamefont {Scuderi}, \citenamefont {Kaufman}, \citenamefont {Wiste},
  \citenamefont {Lastovicka}, \citenamefont {Picciotto}, \citenamefont
  {Bagolini}, \citenamefont {Crivellari}, \citenamefont {Bellutti},
  \citenamefont {Milluzzo}, \citenamefont {Cirrone}, \citenamefont {Magnusson},
  \citenamefont {Gonoskov}, \citenamefont {Korn}, \citenamefont {Wahlstr\"om},\
  and\ \citenamefont {Margarone}}]{giuffrida.prab.2017}%
  \BibitemOpen
  \bibfield  {author} {\bibinfo {author} {\bibfnamefont {L.}~\bibnamefont
  {Giuffrida}}, \bibinfo {author} {\bibfnamefont {K.}~\bibnamefont {Svensson}},
  \bibinfo {author} {\bibfnamefont {J.}~\bibnamefont {Psikal}}, \bibinfo
  {author} {\bibfnamefont {M.}~\bibnamefont {Dalui}}, \bibinfo {author}
  {\bibfnamefont {H.}~\bibnamefont {Ekerfelt}}, \bibinfo {author}
  {\bibfnamefont {I.}~\bibnamefont {Gallardo~Gonzalez}}, \bibinfo {author}
  {\bibfnamefont {O.}~\bibnamefont {Lundh}}, \bibinfo {author} {\bibfnamefont
  {A.}~\bibnamefont {Persson}}, \bibinfo {author} {\bibfnamefont
  {P.}~\bibnamefont {Lutoslawski}}, \bibinfo {author} {\bibfnamefont
  {V.}~\bibnamefont {Scuderi}}, \bibinfo {author} {\bibfnamefont
  {J.}~\bibnamefont {Kaufman}}, \bibinfo {author} {\bibfnamefont
  {T.}~\bibnamefont {Wiste}}, \bibinfo {author} {\bibfnamefont
  {T.}~\bibnamefont {Lastovicka}}, \bibinfo {author} {\bibfnamefont
  {A.}~\bibnamefont {Picciotto}}, \bibinfo {author} {\bibfnamefont
  {A.}~\bibnamefont {Bagolini}}, \bibinfo {author} {\bibfnamefont
  {M.}~\bibnamefont {Crivellari}}, \bibinfo {author} {\bibfnamefont
  {P.}~\bibnamefont {Bellutti}}, \bibinfo {author} {\bibfnamefont
  {G.}~\bibnamefont {Milluzzo}}, \bibinfo {author} {\bibfnamefont {G.~A.~P.}\
  \bibnamefont {Cirrone}}, \bibinfo {author} {\bibfnamefont {J.}~\bibnamefont
  {Magnusson}}, \bibinfo {author} {\bibfnamefont {A.}~\bibnamefont {Gonoskov}},
  \bibinfo {author} {\bibfnamefont {G.}~\bibnamefont {Korn}}, \bibinfo {author}
  {\bibfnamefont {C.-G.}\ \bibnamefont {Wahlstr\"om}}, \ and\ \bibinfo {author}
  {\bibfnamefont {D.}~\bibnamefont {Margarone}},\ }\href {\doibase
  10.1103/PhysRevAccelBeams.20.081301} {\bibfield  {journal} {\bibinfo
  {journal} {Phys. Rev. Accel. Beams}\ }\textbf {\bibinfo {volume} {20}},\
  \bibinfo {pages} {081301} (\bibinfo {year} {2017})}\BibitemShut {NoStop}%
\bibitem [{\citenamefont {Mackenroth}\ \emph {et~al.}(2016)\citenamefont
  {Mackenroth}, \citenamefont {Gonoskov},\ and\ \citenamefont
  {Marklund}}]{Mackenroth}%
  \BibitemOpen
  \bibfield  {author} {\bibinfo {author} {\bibfnamefont {F.}~\bibnamefont
  {Mackenroth}}, \bibinfo {author} {\bibfnamefont {A.}~\bibnamefont
  {Gonoskov}}, \ and\ \bibinfo {author} {\bibfnamefont {M.}~\bibnamefont
  {Marklund}},\ }\href {\doibase 10.1103/PhysRevLett.117.104801} {\bibfield
  {journal} {\bibinfo  {journal} {Phys. Rev. Lett.}\ }\textbf {\bibinfo
  {volume} {117}},\ \bibinfo {pages} {104801} (\bibinfo {year}
  {2016})}\BibitemShut {NoStop}%
\bibitem [{\citenamefont {Wan}\ \emph {et~al.}(2017)\citenamefont {Wan},
  \citenamefont {Zhang}, \citenamefont {Li}, \citenamefont {Wu}, \citenamefont
  {Hua}, \citenamefont {Pai}, \citenamefont {Lu}, \citenamefont {Gu},
  \citenamefont {Joshi},\ and\ \citenamefont {Mori}}]{Wan}%
  \BibitemOpen
  \bibfield  {author} {\bibinfo {author} {\bibfnamefont {Y.}~\bibnamefont
  {Wan}}, \bibinfo {author} {\bibfnamefont {C.~J.}\ \bibnamefont {Zhang}},
  \bibinfo {author} {\bibfnamefont {F.}~\bibnamefont {Li}}, \bibinfo {author}
  {\bibfnamefont {Y.~P.}\ \bibnamefont {Wu}}, \bibinfo {author} {\bibfnamefont
  {J.~F.}\ \bibnamefont {Hua}}, \bibinfo {author} {\bibfnamefont {C.~H.}\
  \bibnamefont {Pai}}, \bibinfo {author} {\bibfnamefont {W.}~\bibnamefont
  {Lu}}, \bibinfo {author} {\bibfnamefont {Y.~Q.}\ \bibnamefont {Gu}}, \bibinfo
  {author} {\bibfnamefont {C.}~\bibnamefont {Joshi}}, \ and\ \bibinfo {author}
  {\bibfnamefont {W.~B.}\ \bibnamefont {Mori}},\ }\href
  {https://arxiv.org/abs/1707.07290} {\bibfield  {journal} {\bibinfo  {journal}
  {arXiv}\ } (\bibinfo {year} {2017})}\BibitemShut {NoStop}%
\bibitem [{\citenamefont {Gonoskov}\ \emph {et~al.}(2009)\citenamefont
  {Gonoskov}, \citenamefont {Korzhimanov}, \citenamefont {Eremin},
  \citenamefont {Kim},\ and\ \citenamefont {Sergeev}}]{gonoskov.prl.2009}%
  \BibitemOpen
  \bibfield  {author} {\bibinfo {author} {\bibfnamefont {A.~A.}\ \bibnamefont
  {Gonoskov}}, \bibinfo {author} {\bibfnamefont {A.~V.}\ \bibnamefont
  {Korzhimanov}}, \bibinfo {author} {\bibfnamefont {V.~I.}\ \bibnamefont
  {Eremin}}, \bibinfo {author} {\bibfnamefont {A.~V.}\ \bibnamefont {Kim}}, \
  and\ \bibinfo {author} {\bibfnamefont {A.~M.}\ \bibnamefont {Sergeev}},\
  }\href {\doibase 10.1103/PhysRevLett.102.184801} {\bibfield  {journal}
  {\bibinfo  {journal} {Phys. Rev. Lett.}\ }\textbf {\bibinfo {volume} {102}},\
  \bibinfo {pages} {184801} (\bibinfo {year} {2009})}\BibitemShut {NoStop}%
\bibitem [{\citenamefont {Bastrakov}\ \emph {et~al.}(2012)\citenamefont
  {Bastrakov}, \citenamefont {Donchenko}, \citenamefont {Gonoskov},
  \citenamefont {Efimenko}, \citenamefont {Malyshev}, \citenamefont {Meyerov},\
  and\ \citenamefont {Surmin}}]{Bastrakov}%
  \BibitemOpen
  \bibfield  {author} {\bibinfo {author} {\bibfnamefont {S.}~\bibnamefont
  {Bastrakov}}, \bibinfo {author} {\bibfnamefont {R.}~\bibnamefont
  {Donchenko}}, \bibinfo {author} {\bibfnamefont {A.}~\bibnamefont {Gonoskov}},
  \bibinfo {author} {\bibfnamefont {E.}~\bibnamefont {Efimenko}}, \bibinfo
  {author} {\bibfnamefont {A.}~\bibnamefont {Malyshev}}, \bibinfo {author}
  {\bibfnamefont {I.}~\bibnamefont {Meyerov}}, \ and\ \bibinfo {author}
  {\bibfnamefont {I.}~\bibnamefont {Surmin}},\ }\href {\doibase
  10.1016/j.jocs.2012.08.012} {\bibfield  {journal} {\bibinfo  {journal} {J.
  Comput. Sci.}\ }\textbf {\bibinfo {volume} {3}},\ \bibinfo {pages} {474}
  (\bibinfo {year} {2012})}\BibitemShut {NoStop}%
\bibitem [{\citenamefont {Surmin}\ \emph {et~al.}(2016)\citenamefont {Surmin},
  \citenamefont {Bastrakov}, \citenamefont {Efimenko}, \citenamefont
  {Gonoskov}, \citenamefont {Korzhimanov},\ and\ \citenamefont
  {Meyerov}}]{Surmin}%
  \BibitemOpen
  \bibfield  {author} {\bibinfo {author} {\bibfnamefont {I.}~\bibnamefont
  {Surmin}}, \bibinfo {author} {\bibfnamefont {S.}~\bibnamefont {Bastrakov}},
  \bibinfo {author} {\bibfnamefont {E.}~\bibnamefont {Efimenko}}, \bibinfo
  {author} {\bibfnamefont {A.}~\bibnamefont {Gonoskov}}, \bibinfo {author}
  {\bibfnamefont {A.}~\bibnamefont {Korzhimanov}}, \ and\ \bibinfo {author}
  {\bibfnamefont {I.}~\bibnamefont {Meyerov}},\ }\href {\doibase
  https://doi.org/10.1016/j.cpc.2016.02.004} {\bibfield  {journal} {\bibinfo
  {journal} {Computer Physics Communications}\ }\textbf {\bibinfo {volume}
  {202}},\ \bibinfo {pages} {204 } (\bibinfo {year} {2016})}\BibitemShut
  {NoStop}%
\end{thebibliography}%


\end{document}